\documentclass[aps,prd,preprint,groupedaddress,showpacs,nofootinbib,citeautoscript,superscriptaddress]{revtex4}
\usepackage{epsfig}
\usepackage{graphics}
\usepackage{amssymb,amsmath}
\usepackage{color}

\begin{document}

\leftmargin -2cm
\def\choosen{\atopwithdelims..}


\preprint{DESY~16-095\hspace{12cm}ISSN~0418-9833}
\preprint{June 2016\hspace{15.18cm}}

\boldmath
\title{$\psi(2S)$ and $\Upsilon(3S)$ hadroproduction in the parton Reggeization
approach: Yield, polarization, and the role of fragmentation}
\unboldmath

\author{\firstname{B.~A.}\surname{Kniehl}}
\email{kniehl@desy.de}
\affiliation{{II.} Institut f\"ur Theoretische Physik, Universit\"at Hamburg,
Luruper Chaussee 149, 22761 Hamburg, Germany}
\author{\firstname{M.~A.}\surname{Nefedov}}
\email{nefedovma@gmail.com}
\author{\firstname{V.~A.}\surname{Saleev}}
\email{saleev@samsu.ru}
\affiliation{{II.} Institut f\"ur Theoretische Physik, Universit\"at Hamburg,
Luruper Chaussee 149, 22761 Hamburg, Germany}
\affiliation{Samara National Research University, Moscow Highway, 34, 443086,
Samara, Russia}

\begin{abstract}
The hadroproduction of the radially excited heavy-quarkonium states $\psi(2S)$
and $\Upsilon(3S)$ at high energies is studied in the parton Reggeization
approach and the factorization formalism of nonrelativistic QCD at lowest
order in the strong-coupling constant $\alpha_s$ and the relative heavy-quark
velocity $v$.
A satisfactory description of the $\psi(2S)$ transverse-momentum ($p_T$)
distributions measured by ATLAS, CMS, and LHCb at center-of-mass energy
$\sqrt{S}=7$~TeV is obtained using the color-octet long-distance matrix
elements (LDMEs) extracted from CDF data at $\sqrt{S}=1.96$~TeV.
The importance of the fragmentation mechanism and the scale evolution of the
fragmentation functions in the upper $p_T$ range, beyond 30~GeV, is
demonstrated.
The $\Upsilon(3S)$ $p_T$ distributions measured by CDF at $\sqrt{S}=1.8$~TeV
and by LHCb at $\sqrt{S}=7$~TeV and forward rapidities are well described
using LDMEs fitted to ATLAS data at $\sqrt{S}=7$~TeV.
Comparisons of polarization measurements by CDF and CMS at large $p_T$ values
with our predictions consolidate the familiar problem in the $\psi(2S)$ case,
but yield reasonable agreement in the $\Upsilon(3S)$ case.
\end{abstract}


\maketitle

\section{Introduction}
\label{sec:intro}

The production of heavy quarkonia at hadron colliders is a unique laboratory
for studies of the interplay between the perturbative treatment of hard
subprocesses and nonperturbative hadronization models.
Thanks to the hierarchy $m_{Q}\gg\Lambda_\mathrm{QCD}$, where $m_Q$ is the mass
of the heavy quark $Q=c,b$ and $\Lambda_\mathrm{QCD}$ is the asymptotic scale
parameter of quantum chromodynamics (QCD), the nonrelativistic-QCD (NRQCD)
factorization hypothesis \cite{NRQCD} (see also the recent reviews in
Ref.~\cite{QWGrev}) allows one to factorize the effects of short and long
distances and to parametrize the latter in terms of a few long-distance matrix
elements (LDMEs).
While color-singlet (CS) LDMEs are calculable in potential models
\cite{EitchenQuigg}, the only way to extract color-octet (CO) LDMEs available
so far is to fit them to experimental data.
This implies that, to reliably check the validity of NRQCD factorization and
the universality of the LDMEs, one has to know the short-distance parts of the
cross sections as precisely as possible.
The hadroproduction of heavy quarkonia is presently being studied in a wide
range of transverse momentum ($p_T$) and both at central and forward rapidities
($y$).
To provide a uniform and accurate description of the short-distance parts of
the cross sections is a challenging task even with state-of-the-art techniques
in perturbative QCD.

Three characteristic $p_T$ regions can be identified.
In the region $p_T\alt M$, where $M$ is the heavy-quarkonium mass, Sudakov-type
double logarithms $\ln^2(p_T/M)$ spoil the convergence of the perturbative
series in $\alpha_s$ and have to be resummed to reproduce the physical behavior
of the cross section \cite{smallPTR}.
Moreover, small-$x$ physics effects, such as the saturation of parton
distribution functions (PDFs), can start to play a role there.
In fact, at $\sqrt{S}=7$~TeV, $x$ values as small as $10^{-5}$ contribute to
the lowest $p_T$ bins for the rapidities covered by the LHCb detector
\cite{HQsatur}.
At $p_T\gg M$, fragmentation logarithms $\ln(p_T/M)$ appear, and the
description in terms of fragmentation functions, evolving with the energy
scale, appears to be more appropriate \cite{Frag_GS_YQM}.
In some intermediate $p_T$ region, fixed-order calculations within the
collinear parton model (CPM) should be valid.
In the CPM, the complete next-to-leading-order (NLO) results for inclusive
heavy-quarkonium production are available
\cite{KniehlPSI,Ma:2010yw}.
The real-radiation part of the next-to-next-to-leading-order corrections to CS
production was found to be sizable \cite{NNLOstar}, even taking into account
the large uncertainties due to the infrared cutoff scale.

The above-mentioned approaches appear to describe well the $p_T$ distributions
measured in the respective regions.
However, there is dramatic disagreement between the CO LDME sets extracted in
different fits.
Moreover, while a self-consistent description of all the experimental data of
prompt $J/\psi$ hadroproduction and photoproduction is possible at NLO in the
CPM \cite{Butenschoen:2009zy,KniehlPSI}, the LDMEs thus obtained lead to
disagreement with the polarization measurements \cite{KBpol}.
A similar, albeit less severe tension between the descriptions of yield and
polarization was also observed for bottomonia \cite{GWZUpsi}.
This problem is usually referred to as the {\it heavy-quarkonium polarization
puzzle}.

In view of the difficult situation described above, an approach which is
equally appropriate on theoretical grounds both for the small- and large-$p_T$
regions is required.
Such an approach can be designed on the basis of the $k_T$ factorization
formalism \cite{kTf} implemented with PDFs unintegrated over $p_T$ (unPDFs),
which naturally regularizes the small-$p_T$ divergences that are present in
fixed-order calculations within the CPM.
The gauge independence of the hard-scattering matrix elements is, in general,
broken by the virtuality of the initial-state gluons.
To restore it, one can treat them as Reggeized gluons (Reggeons), which are the
natural gauge-independent degrees of freedom of high-energy QCD.
They were first introduced in the context of the
Balitsky-Fadin-Kuraev-Lipatov (BFKL) \cite{BFKL} evolution equation and
later promoted to the level of dynamical fields in Lipatov's effective action
for the high-energy limit of QCD \cite{LipatovEFT}.
We denote the combination of the $k_T$ factorization formalism for the cross
sections with the Reggeization of partons in the initial state of the
hard-scattering amplitudes as the parton Reggeization approach (PRA).

Presently, unPDFs are not so much constrained as collinear PDFs.
However, there exists a method to obtain unPDFs from collinear ones, the
Kimber-Martin-Ryskin (KMR) \cite{KMR} model, which has produced stable and
consistent results in many phenomenological applications.
Besides numerous applications to charmonium
\cite{Frag_YaF,KSVcharm,Frag_PRD,SVYadFiz,SNScharm} and bottomonium production
\cite{SVYadFiz,KSVbottom,SNSbottom}, the PRA with KMR unPDFs has recently been
successfully applied to describe the production of open charm \cite{OpenCharm},
$B$ mesons \cite{Bmesons}, dijets \cite{dijets}, bottom-flavored jets
\cite{bjets}, Drell-Yan lepton pairs \cite{DY}, monojets, and prompt photons
\cite{PPSJ} at the Fermilab Tevatron and the CERN LHC and to the associated
production of photons and jets at DESY HERA
\cite{PPJHera}.

In the present paper, we concentrate on the production of radially excited
charmonium [$\psi(2S)$] and bottomonium [$\Upsilon(3S)$] states.
This has the advantage that the feed-down contributions are negligibly small
and so allows for direct tests of the underlying production mechanisms.
The recent experimental data on the unpolarized $\psi(2S)$ yields from ATLAS
\cite{ATLAS2014} and CMS \cite{CMS2015} cover a wide $p_T$ range and, in
combination with CDF \cite{CDF2009} and LHCb \cite{LHCb2012} data at smaller
$p_T$ values, allow us to quantitatively study the relative importance of the
fusion and fragmentation production mechanisms.
Measurements of $J/\psi$ production from $\psi(2S)$ decay by CDF \cite{CDF1997}
and ATLAS \cite{ATLAS2014} enable us to test a simple model of the feed-down
kinematics \cite{GWZUpsi,SHMZCpsi}.
Furthermore, we exploit $\psi(2S)$ polarization data from CDF
\cite{CDFpol_psi2S} and CMS \cite{CMSpol} to address the question if the PRA
can shed light on the notorious charmonium polarization puzzle.
In the $\Upsilon(3S)$ case, we apply the PRA to interpret unpolarized-yield
data by CDF \cite{Upsi_CDF2002}, ATLAS \cite{Upsi_ATLAS2013}, and LHCb
\cite{Upsi_LHCb2012} and polarization data by CDF \cite{CDFpol} and CMS
\cite{Chatrchyan:2012woa}.

This paper is organized as follows.
In Sec.~\ref{sec:basicPRA}, we outline the basics of the PRA.
Specifically, we describe both the fusion and fragmentation approximations at
leading order (LO) in Sec.~\ref{sec:unpol}, and we list our analytic results
for the polarization observables in Sec.~\ref{sec:pol}.
In Sec.~\ref{sec:spectra}, we compare the selected experimental data with our
numerical results.
Specifically, Sec.~\ref{sec:Unpolar} is devoted to the unpolarized yields and 
Sec.~\ref{sec:Polar} to the polarization observables.
In Sec.~\ref{sec:Concl}, we interpret the obtained results and summarize our
conclusions.

\section{Basic formalism}
\label{sec:basicPRA}

\subsection{Unpolarized yields}
\label{sec:unpol}

The NRQCD factorization formalism \cite{NRQCD} suggests that the effects of
short and long distances are factorized in the partonic cross sections of the
production of the heavy-quarkonium state ${\cal H}$ as
\begin{equation}
d\hat{\sigma}^{\cal H}=\sum\limits_n d\hat\sigma(Q\bar{Q}[n])
\langle{\cal O}^{\cal H}[n]\rangle,
\label{eq:NRQCDFT}
\end{equation}
where the sum is over the possible intermediate Fock states
$n={}^{2S+1}\!L_J^{(a)}$ of the $Q\bar{Q}$ pair, with definite spin $S$, orbital
momentum $L$, total angular momentum $J$, and CS or CO quantum numbers
$a=1,8$, respectively.
The decomposition in Eq.~(\ref{eq:NRQCDFT}) corresponds to a double expansion
in the strong-coupling constant $\alpha_s$ and the relative heavy-quark
velocity $v$.
The short-distance cross sections $d\hat\sigma (Q\bar{Q}[n])$ are
perturbatively calculable, and the LDMEs $\langle{\cal O}^{\cal H}[n]\rangle$
possess definite $v$ scaling properties \cite{NRQCDvSR}.
For ${\cal H}=\psi(2S),\Upsilon(3S)$, the CS LDME
$\langle{\cal O^H}[{}^3\!S_1^{(1)}]\rangle$ contributes at ${\cal O}(v^3)$, and
the CO LDMEs
$\langle{\cal O^H}[{}^1\!S_0^{(8)}]\rangle$,
$\langle{\cal O^H}[{}^3\!S_1^{(8)}]\rangle$, and
$\langle{\cal O^H}[{}^3\!P_J^{(8)}]\rangle$ ($J=0,1,2$) contribute at
${\cal O}(v^7)$, while contributions of higher orders in $v$ are usually
neglected.

The dominant contribution to inclusive heavy-quarkonium production at hadron
colliders comes from the gluon fusion subprocess.
In the PRA, its cross section can be represented as
\begin{equation}
d\sigma(pp\to{\cal H}+X)=
\int\frac{dx_1}{x_1}\int\frac{d^2{\bf q}_{T1}}{\pi}\Phi_g(x_1,t_1,\mu_F^2)
\int\frac{dx_2}{x_2}\int\frac{d^2{\bf q}_{T2}}{\pi}\Phi_g(x_2,t_2,\mu_F^2)
d\hat{\sigma}^{\cal H},
\end{equation}
where the four-momenta $q_i$ ($i=1,2$) of the Reggeons are parametrized as
sums of longitudinal and transverse parts, $q_i=x_iP_i+q_{Ti}$, where $P_i$ are
the four-momenta of the colliding protons and $q_{Ti}=(0,{\bf q}_{Ti},0)$.
We have $q_i^2=-{\bf q}_{Ti}^2=-t_i$ and $2P_1\cdot P_2=S$.
In our approach, the gluon unPDF $\Phi_g(x,{\bf q}_T^2,\mu_F^2)$ is normalized
relative to the collinear PDF by the following condition:
\begin{equation}
\int\limits^{\mu_F^2} dt \Phi_g(x,t,\mu_F^2) = x f_g(x,\mu_F^2).
\end{equation}

For the inelastic scattering of objects with hard intrinsic scales, such as
photons with finite virtualities ($Q^2$), at high center-of-mass energies
$\sqrt{S}$, the evolution of the unPDFs is governed by the large logarithms
$\ln(S/Q^2)$ or $\ln(1/x)$ and is subject to the BFKL evolution equation
\cite{BFKL}.
In the production of particles with large $p_T$ values,
$\Lambda_\mathrm{QCD}\ll p_T\ll \sqrt{S}$, in proton-proton collisions, the
initial state does not provide a sufficiently hard intrinsic scale, so that
the $k_T$-ordered Dokshitzer-Gribov-Lipatov-Altarelli-Parisi (DGLAP)
\cite{DGLAP} evolution at small values of $k_T$ should be merged with the
rapidity-ordered BFKL evolution at the final large-$k_T$ steps of the
initial-state-radiation cascade.
The latter problem is highly nontrivial and equivalent to the complete
resummation of the $\ln k_T$-enhanced terms in the BFKL kernel.
A few phenomenological schemes to compute unPDFs of the proton were proposed.
In the present paper, we use the LO KMR unPDFs \cite{KMR}, generated from the
LO set of Martin-Stirling-Thorne-Watt collinear PDFs \cite{MSTW_2008}.
Furthermore, we use the LO formula for $\alpha_s$ with normalization
$\alpha_s(M_Z)=0.12609$ and flavor thresholds at $m_c=1.4$~GeV and
$m_b=4.75$~GeV.

We take into account the following $2\to 1$ and $2\to 2$ partonic
subprocesses:
\begin{eqnarray}
R(q_1)+R(q_2)& \to & Q\bar{Q}\left[{}^1\!S_0^{(8)},{}^3\!S_1^{(8)},
{}^3\!P_J^{(8)}\right],
\nonumber\\
R(q_1)+R(q_2)& \to & Q\bar{Q}\left[{}^3\!S_1^{(1)} \right] + g,
\label{eq:3S11pr}
 \end{eqnarray}
where $R$ denotes the Reggeon.
The matrix elements of the subprocesses in Eq.~(\ref{eq:3S11pr}), summed over
the polarizations of the final-state $Q\bar{Q}$ pair, were obtained in
Ref.~\cite{KSVcharm}.
As shown in Ref.~\cite{KSVcharm}, our normalization conventions for the LDMEs
coincide with those of Ref.~\cite{NRQCDMaltoni}.

In Ref.~\cite{SNScharm}, CO LDMEs were fitted to Tevatron data of prompt
$J/\psi$ production in the following approximation.
The charm-quark mass $m_c$ was taken to be $m_c=M_{J/\psi}/2\approx1.5$~GeV, and
the mass differences between the $J/\psi$ meson and the excited $\chi_{cJ}$ and
$\psi(2S)$ states were neglected in the respective feed-down contributions.
This approximation is consistent with the NRQCD calculation at fixed order in
$v$, since the mass difference $\Delta M$ is proportional to $v^2$ in the
potential models.
However, the kinematic effect of the mass splittings between charmonium
states turns out to be significant.
For the decay ${\cal H}_1\to {\cal H}_2+X$, the following approximate relation
between the transverse momenta is valid in the limit
$\Delta M\ll M_{{\cal H}_{1,2}}$:
 \begin{equation}
\langle p_T^{{\cal H}_2} \rangle
= \frac{M_{{\cal H}_2}}{M_{{\cal H}_1}} p_T^{{\cal H}_1} 
+ {\cal O}\left(\frac{(\Delta M)^2}{M^2},\frac{M}{p_T}\right),
\label{eq:pT_shift}
\end{equation}
where the averaging on the left-hand side is performed over the uniform
distribution of the decay products in the rest frame of ${\cal H}_1$.
Due to the powerlike decrease of the $p_T$ distribution at large $p_T$ values,
the small $p_T$ shift in Eq.~(\ref{eq:pT_shift}) can lead to a change in cross
section by up to a factor of 2 in the case of charmonia and by up to 20\%--30\%
in the case of bottomonia.
In LO NRQCD calculations, the mass splitting can be taken into account only by
appropriately adjusting the quark mass.
In the present paper, we thus take $m_{c,b}=M_{\cal H}/2$.
This approximation together with the shift in Eq.~(\ref{eq:pT_shift}) was
first adopted in Refs.~\cite{GWZUpsi,SHMZCpsi}.
We would like to stress that the use of this kind of kinematic approximations
actually violates the fixed-order character of the expansion in $v$ implied by
Eq.~(\ref{eq:NRQCDFT}).

Since the LHC data cover values of $p_T$ all the way up to 100~GeV,
fragmentation corrections may be of vital importance for their description.
In the LO-in-$\alpha_s$ plus leading-logarithmic (LL) approximation, only the
$g\to Q\bar{Q}[{}^3\!S_1^{(8)}]$ transition acquires large logarithmic
corrections of the type $\alpha_s\ln(p_T/M)$.
In the large-$p_T$ regime, the cross section of $pp\to{\cal H}+X$ may thus be
approximately calculated as
\begin{equation}
\frac{d\sigma}{dp_T^{\cal H}dy_{\cal H}}(pp\to{\cal H}+X)
=\int\limits_0^1 dz
\frac{d\sigma}{dp_T^gdy_g}(pp\to g)
D_{g\to {\cal H}\left[{}^3\!S_1^{(8)}\right]}(z,\mu_F^2),
\label{eq:FRAG}
\end{equation}
where $p_T^g=p_T^{\cal H}/z$ and $y_g=y_{\cal H}$.
To LO in the PRA, we have
\begin{equation}
\frac{d\sigma}{dp_T^g dy_g}(pp\to g)=\frac{1}{(p_T^g)^3}
\int\limits_0^\infty dt_1\int\limits_0^{2\pi} d\phi_1
\Phi_g(x_1,t_1,\mu_F^2) \Phi_g(x_2,t_2,\mu_F^2)
\overline{|{\cal M}(RR\to g)|^2},
\end{equation}
where $\overline{|{\cal M}(RR\to g)|^2}=(3/2)\pi\alpha_s(\mu_R^2)(p_T^g)^2$ is
the squared amplitude obtained from the Fadin-Lipatov effective
Reggeon-Reggeon-gluon vertex \cite{BFKL,PPSJ} and
$t_2=t_1+(p_T^g)^2-2p_T^g\sqrt{t_1}\cos\phi_1$.
The fragmentation function $D_{g\to {\cal H}[{}^3\!S_1^{(8)}]}(z,\mu_F^2)$ is obtained
by solving the LO DGLAP evolution equation \cite{DGLAP} with the initial
condition
\begin{equation}
D_{g\to {\cal H}\left[{}^3\!S_1^{(8)}\right]}(z,\mu_{F0}^2)
=\frac{\pi\alpha_s(\mu_{F0}^2)}{6M_{\cal H}^3}
\left\langle {\cal O^H}\left[{}^3\!S_1^{(8)}\right]\right\rangle\delta(1-z),
\label{eq:FF0}
\end{equation}
at the starting scale $\mu_{F0}^2=M_{\cal H}^2$.
The explicit form of the solution can be found, {\it e.g.}, in
Ref.~\cite{Frag_YaF}.
In the following, we shall refer to the production mechanism underlying
Eqs.~(\ref{eq:FRAG})--(\ref{eq:FF0}) as fragmentation and the one
underlying the usual treatment of the ${}^3\!S_1^{(8)}$ contribution 
\cite{KSVcharm,Frag_PRD,SVYadFiz,SNScharm,KSVbottom,SNSbottom} as fusion.

We take the renormalization and factorization scales to be
$\mu_F=\mu_R=\xi M_T$, where $M_T=\sqrt{M_{\cal H}^2+p_T^2}$ is the transverse
mass, and vary $\xi$ by a factor of 2 up and down about the default value 1 to
estimate the scale uncertainty.

\subsection{Polarization parameters}
\label{sec:pol}

The polarization parameters $\lambda_\theta$, $\lambda_\varphi$, and
$\lambda_{\theta\varphi}$ are defined through the angular distribution of the
leptonic decay ${\cal H}\to l^+l^-$ in the rest frame of the ${}^3\!S_1$
heavy-quarkonium state ${\cal H}$,
  \begin{equation}
  \frac{d\sigma}{d\Omega}\propto 1+\lambda_\theta \cos^2\theta
+\lambda_\varphi \sin^2\theta \cos(2\varphi)
   + \lambda_{\theta\varphi} \sin(2\theta) \cos\varphi,
  \end{equation}
where $\theta$ and $\varphi$ are the polar and azimuthal angles of the flight
direction of lepton $l^+$ in some appropriately chosen coordinate system.
This choice is an important issue, which is widely discussed in the literature,
see, {\it e.g.}, Refs.~\cite{POLfrms,Faccioli:2010kd}.
In the present study, we concentrate on the polarization parameter
$\lambda_\theta$ in the $s$-channel helicity frame, where the
longitudinal polarization vector points along the $z$ axes and can be written
in covariant form as
   \begin{equation}
   \varepsilon^\mu(0)=Z^\mu
=\frac{(p\cdot Q)p^\mu/M-MQ^\mu}
{\sqrt{(p\cdot Q)^2-M^2 S}},
   \end{equation}
with $p$ being the four-momentum of ${\cal H}$ and $Q=P_1+P_2$.
The calculation of $\lambda_\theta$ in the PRA proceeds along the same lines as
in the CPM \cite{POLfrms,BKL,KniehlLee}, and we merely list our results.
We have
\begin{equation}
\lambda_\theta=\frac{\sigma^{\cal H}-3\sigma^{\cal H}_L}
{\sigma^{\cal H}+\sigma^{\cal H}_L},
\label{eq:Ltheta}
\end{equation}
where $\sigma_L^{\cal H}$ is the cross section for the production of the
heavy-quarkonium state ${\cal H}$ with longitudinal polarization, $J_z=0$, and
$\sigma^{\cal H}$ is summed over $J_z=0,\pm1$ as in Sec.~\ref{sec:unpol}.
Assuming the polarizations of the ${}^3\!S_1^{(1)}$ and ${}^3\!S_1^{(8)}$ states
to be directly transferred to the $\mathcal{H}$ meson and chromoelectric-dipole
transitions with $\Delta S=0$ and $\Delta L= 1$ from the ${}^3\!P_J$ states
\cite{PolTransfer}, we have
 \begin{eqnarray}
\sigma_L^{\cal H}&=&\sigma_0^{\cal H}\left[{}^3\!S_1^{(1)}\right]
+\sigma_0^{\cal H}\left[{}^3\!S_1^{(8)}\right]
+\frac{1}{3}\left(\sigma^{\cal H}\left[{}^1\!S_0^{(8)}\right]
+\sigma^{\cal H}\left[{}^3\!P_0^{(8)}\right]\right)
+\frac{1}{2}\left(\sigma_1^{\cal H}\left[{}^3\!P_1^{(8)}\right]
+\sigma_1^{\cal H} \left[{}^3\!P_2^{(8)}\right]\right)
\nonumber\\
&&{}
+\frac{2}{3}\sigma_0^{\cal H}\left[{}^3\!P_2^{(8)}\right],
\end{eqnarray}
where the label $J_z$ in the notation $\sigma_{|J_z|}^{\cal H}[n]$ refers to the
$Q\bar{Q}$ Fock state $n$ rather than the heavy-quarkonium state $\mathcal{H}$.
The relevant matrix element squares
$\overline{\vert{\cal M}_{|J_z|}(RR\to{\cal H}[n])\vert^2}$ for fixed value of
$|J_z|$, averaged over the spins and colors of the incoming Reggeons, are given
by
\begin{eqnarray}
\overline{\left\vert{\cal M}_0\left(RR\to{\cal H}\left[{}^3\!S_1^{(8)}\right]
\right)\right\vert^2}
&=&2\pi^2\alpha_s^2\frac{\left\langle{\cal O}^{\cal H}\left[{}^3\!S_1^{(8)}\right]
\right\rangle}{M^3}
\nonumber\\
&&{}\times\frac{M^2M_T^2(t_1x_1-t_2x_2)^2\cos^2\varphi}
{(M^2+t_1+t_2)^2\left[M^2(x_1-x_2)^2+p_T^2(x_1+x_2)^2\right]},
\nonumber\\
\overline{\left\vert{\cal M}_0\left(RR\to{\cal H}\left[{}^3\!P_1^{(8)}\right]
\right)\right\vert^2}
&=&\overline{\left\vert{\cal M}\left(RR\to{\cal H}\left[{}^3\!P_1^{(8)}\right]
\right)\right\vert^2}
-\overline{\left\vert{\cal M}_1\left(RR\to{\cal H}\left[{}^3\!P_1^{(8)}\right]
\right)\right\vert^2}
\nonumber\\
&=&\frac{20}{9}\pi^2\alpha_s^2
\frac{\left\langle{\cal O}^{\cal H}\left[{}^3\!P_1^{(8)}\right]\right\rangle}{M^5}
\nonumber\\
&&{}\times\frac{M^2M_T^6(t_1+t_2)^2(x_1+x_2)^2\sin^2\varphi}
{(M^2+t_1+t_2)^4\left[M^2(x_1-x_2)^2+p_T^2(x_1+x_2)^2\right]},
\nonumber\\
\overline{\left\vert{\cal M}_0\left(RR\to{\cal H}\left[{}^3\!P_2^{(8)}\right]
\right)\right\vert^2}
&=&\frac{4}{9}\pi^2\alpha_s^2
\frac{\left\langle{\cal O}^{\cal H}\left[{}^3\!P_2^{(8)}\right]\right\rangle}{M^5}
\nonumber\\
&&{}\times\frac{M^2M_T^4\left[M_T^2(x_1+x_2)^2+2M^2x_1x_2\right]^2}
{(M^2+t_1+t_2)^4\left[M^2(x_1-x_2)^2+p_T^2(x_1+x_2)^2\right]^2}
\nonumber\\
&&{}\times\left[(t_1+t_2)\cos\varphi+2\sqrt{t_1t_2}\,\right]^2,
\nonumber\\
\overline{\left\vert{\cal M}_1\left(RR\to{\cal H}\left[{}^3\!P_2^{(8)}\right]
\right)\right\vert^2}
&=&\frac{4}{3}\pi^2\alpha_s^2
\frac{\left\langle{\cal O}^{\cal H}\left[{}^3\!P_2^{(8)}\right]\right\rangle}{M^5}
\nonumber\\
&&{}\times\frac{M^4M_T^6(x_1+x_2)^2}
{(M^2+t_1+t_2)^4\left[M^2(x_1-x_2)^2+p_T^2(x_1+x_2)^2\right]^2}
\nonumber\\
&&{}\times
\left\{p_T^2\left[M_T^2(x_1^2+x_2^2)-2M^2x_1x_2\right]
-2x_1x_2\left[(t_1^2+t_2^2)\cos(2\varphi)
\right.\right.
\nonumber\\
&&{}+\left.\left.
4(t_1+t_2)\sqrt{t_1t_2}\cos\varphi
+6t_1t_2\right]\right\},
\end{eqnarray}
where $\varphi$ is the angle enclosed between ${\bf q}_{T1}$ and ${\bf q}_{T2}$.
Our result for a longitudinally polarized ${}^3\!S_1^{(1)}$ state is too lengthy
to be present here.

\section{Numerical results}
\label{sec:spectra}

\subsection{Unpolarized yields}
\label{sec:Unpolar}

We are now in a position to compare the $p_T$ distributions of unpolarized
$\psi(2S)$ and $\Upsilon(3S)$ mesons measured at the Tevatron and the LHC
with our theoretical predictions.
The values of the CS LDMEs
$\left\langle{\cal O}^{\psi(2S)}\left[{}^3\!S_1^{(1)}\right]\right\rangle$ and
$\left\langle{\cal O}^{\Upsilon(3S)}\left[{}^3\!S_1^{(1)}\right]\right\rangle$
listed in Table~\ref{tab:NMEs} are adopted from Refs.~\cite{BKL,EitchenQuigg},
where they were determined from the total width of the $\psi(2S)\to\mu^+\mu^-$
decay and a potential model, respectively.

\begin{figure}
\begin{center}
\includegraphics[width=0.9\textwidth]{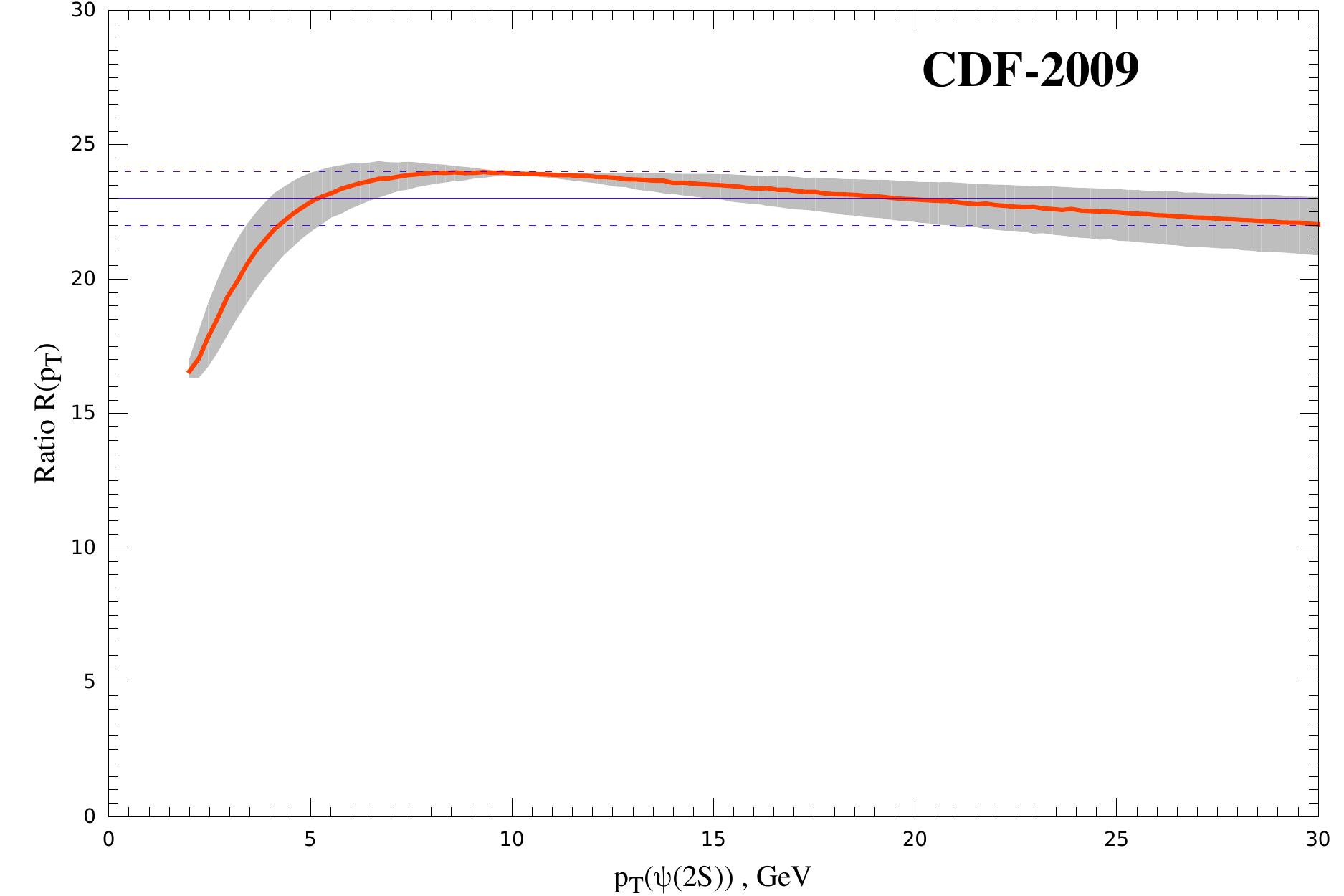}
\end{center}
\caption{\label{fig:ratio}%
Ratio $R_{\psi(2S)}$ defined in Eq.~(\ref{eq:R_def}) as a function on $p_T$
under CDF-2009 \cite{CDF2009} kinematic conditions (thick solid orange line)
and its theoretical uncertainty (shaded band).
The average value $R_{\psi(2S)}=23.0\pm 1.0$ (thin solid and dashed blue lines)
is shown for comparison.}
\end{figure}

\begin{figure}
\begin{center}
\includegraphics[width=0.9\textwidth]{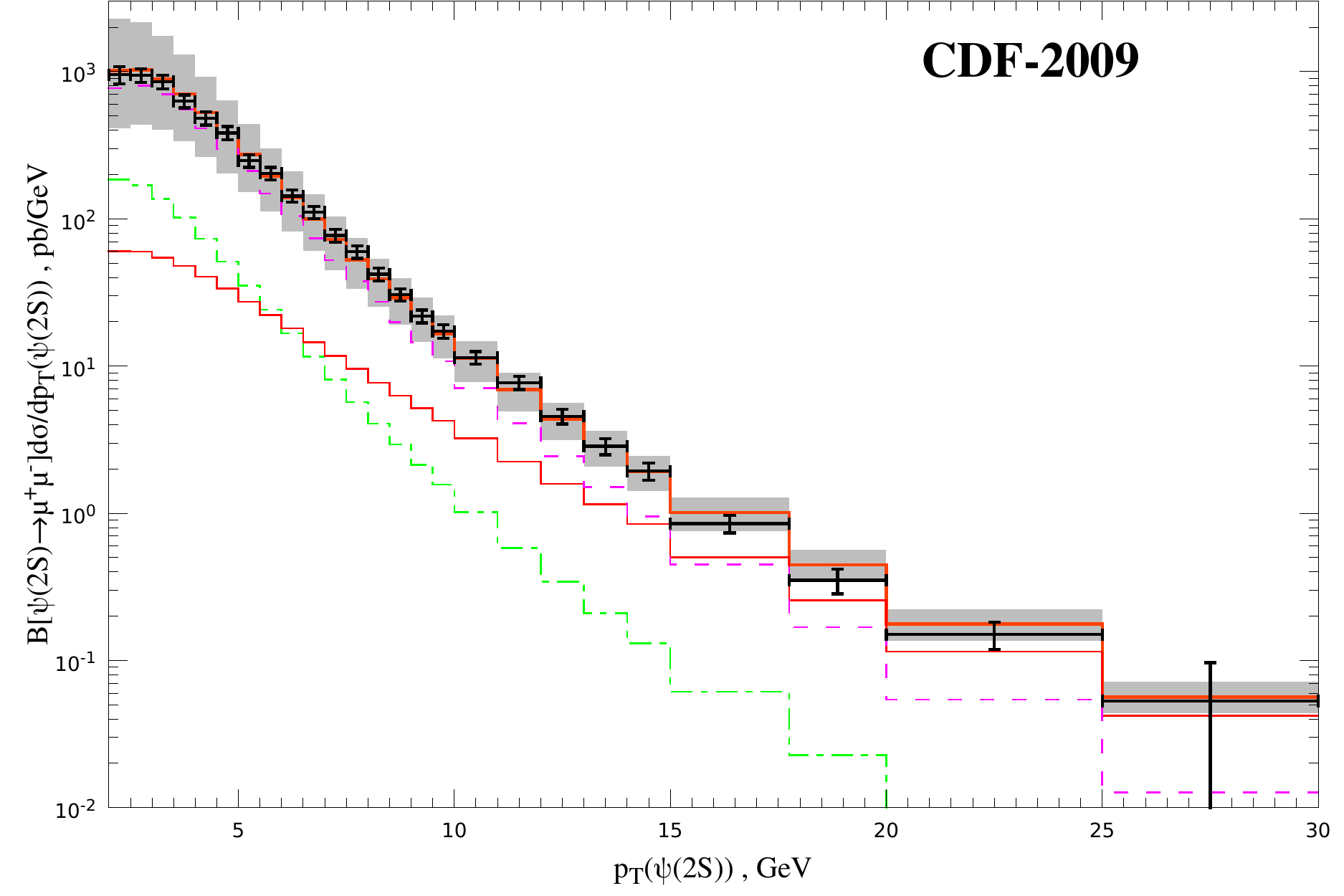}
\end{center}
\caption{\label{fig:CDF2009}%
The CDF-2009 \cite{CDF2009} data set on the $p_T$ distribution of $\psi(2S)$
inclusive hadroproduction is compared with the fitted LO PRA result in the
fusion approximation (thick solid orange histogram) and its theoretical
uncertainty (shaded band).
The ${}^3\!S_1^{(1)}$ (thin dot-dashed green histogram),
${}^3\!S_1^{(8)}$ (thin solid red histogram), and
mixed ${}^1\!S_0^{(8)}$ and ${}^3\!P_J^{(8)}$ (thin dashed violet histogram)
contributions are shown for comparison.}
\end{figure}

We start with the $\psi(2S)$ case.
The CDF Collaboration measured the $p_T$ distribution of prompt $\psi(2S)$
mesons at $\sqrt{S}=1.96~\mbox{TeV}$ for pseudorapidities $|\eta|<0.6$ in the
range $2~\mbox{GeV}<p_T<30~\mbox{GeV}$ by reconstructing their
$\psi(2S)\to\mu^+\mu^-$ decays (CDF-2009) \cite{CDF2009}.
Here and in the following, $p_T\equiv p_T^{\psi(2S)}$, $y\equiv y_{\psi(2S)}$, and
$\eta\equiv\eta_{\psi(2S)}$.
For such moderate $p_T$ values, the fusion approximation is expected to be
appropriate.
At LO and NLO in the CPM, fits of the $J/\psi$ and $\psi(2S)$ CO LDMEs to
hadroproduction data are known to fail to separately determine
$\left\langle{\cal O^H}\left[{}^1\!S_0^{(8)}\right]\right\rangle$ and
$\left\langle{\cal O^H}\left[{}^3\!P_J^{(8)}\right]\right\rangle$ because the
respective $p_T$ distributions exhibit very similar line shapes
\cite{KniehlPSI,SHMZCpsi}.
In Fig.~\ref{fig:ratio}, we investigate if this problem carries over to the
PRA by considering the ratio
\begin{equation}
R_{\cal H}(p_T)=\frac{M_{\cal H}^2
\sum\limits_{J=0}^2(2J+1)d\sigma/dp_T\left[{}^3\!P_J^{(8)}\right]}
{d\sigma/dp_T\left[{}^1\!S_0^{(8)}\right]},
\label{eq:R_def}
\end{equation}
for ${\cal H}=\psi(2S)$ together with its scale uncertainty as a function of
$p_T$ under CDF-2009 kinematic conditions.
We observe that the fraction $R_{\psi(2S)}(p_T)$ varies very feebly in the
interval $5~\mbox{GeV}<p_T<30~\mbox{GeV}$ and can be well approximated by the
constant $R_{\psi(2S)}=23.0\pm1.0$, while its numerator and denominator
themselves vary by several orders of magnitude.
In view of the considerable experimental errors and the scale uncertainties of
the theoretical predictions, it is thus unfeasible to separately determine
$\left\langle{\cal O}^{\psi(2S)}\left[{}^1\!S_0^{(8)}\right]\right\rangle$ and
$\left\langle{\cal O}^{\psi(2S)}\left[{}^3\!P_0^{(8)}\right]\right\rangle$ by
just fitting large-$p_T$ data.
Instead, we introduce the linear combination
\begin{equation}
M_R^{\cal H}=
\left\langle{\cal O}^{\cal H}\left[{}^1\!S_0^{(8)}\right]\right\rangle
+\frac{R_{\cal H}}{M_{\cal H}^2}
\left\langle {\cal O}^{\cal H}\left[{}^3\!P_0^{(8)}\right]\right\rangle
\label{eq:m0}
\end{equation}
for ${\cal H}=\psi(2S)$.
Our fit to the CDF-2009 \cite{CDF2009} data is excellent, as is evident from
Fig.~\ref{fig:CDF2009}, yielding just $\chi^2/\mbox{d.o.f.}=0.6$.
The resulting fit parameters are listed in Table~\ref{tab:NMEs}.

\begin{turnpage}
\begin{table}
\begin{ruledtabular}
\begin{tabular}{c|cccc}
LDME & Fusion & Fragmentation & NLO CPM \cite{GWZUpsi,SHMZCpsi} &
NLO CPM \cite{private} \\
\hline
$\left\langle{\cal O}^{\psi(2S)}\left[{}^3\!S_1^{(1)}\right]\right\rangle/
\mbox{GeV}^3$ & $0.65\pm 0.06$ \cite{BKL} & $0.65\pm 0.06$ \cite{BKL} &
0.76 \cite{EitchenQuigg} & 0.76 \cite{EitchenQuigg} \\
$\left\langle{\cal O}^{\psi(2S)}\left[{}^3\!S_1^{(8)}\right]\right\rangle/
\mbox{GeV}^3\times10^3$ & 
$1.84\pm 0.23$ & $2.57\pm 0.09$ & $1.2 \pm 0.3$ & $2.80\pm 0.49$ \\
$M_R^{\psi(2S)}/\mbox{GeV}^3\times10^2$ &
$3.11\pm0.14$ & $2.70\pm0.11$ & $2.0\pm 0.6$ & $0.37\pm4.85$ \\
$R_{\psi(2S)}$ & $23.0\pm1.0$ & $23.0\pm 1.0$ & $23.5$ & $23.0$ \\
\hline
$\chi^2/\mbox{d.o.f.}$ & 0.6 & 1.1 & 0.56 & 2.84 \\
\hline
$\left\langle{\cal O}^{\Upsilon(3S)}\left[{}^3\!S_1^{(1)}\right]\right\rangle/
\mbox{GeV}^3$ &
3.54 \cite{EitchenQuigg} & $\cdots$ & 3.54 \cite{EitchenQuigg} & $\cdots$ \\
$\left\langle{\cal O}^{\Upsilon(3S)}\left[{}^3\!S_1^{(8)}\right]\right\rangle/
\mbox{GeV}^3\times10^2$ &
$2.73\pm 0.15$ & $\cdots$ & $2.71\pm 0.13$ & $\cdots$ \\
$M_R^{\Upsilon(3S)}/\mbox{GeV}^3\times 10^2$ &
$0.00\pm 0.18$ & $\cdots$ & $1.08\pm 1.66$ & $\cdots$ \\
$R_{\Upsilon(3S)}$ & $22.1\pm0.7$ & $\cdots$ & $22.1$ & $\cdots$ \\
\hline
$\chi^2/\mbox{d.o.f.}$ & 9.7 & $\cdots$ & 3.16 & $\cdots$
\end{tabular}
\end{ruledtabular}
\caption{$\psi(2S)$ and $\Upsilon(3S)$ LDME sets.
The CS LDMEs \cite{EitchenQuigg,BKL} are input.
The $\psi(2S)$ CO LDMEs are fitted to the CDF-2009 \cite{CDF2009} data in the
fusion approximation and to the ATLAS-2014 \cite{ATLAS2014} and the CMS-2015
\cite{CMS2015} data in the fragmentation approximation.
The $\Upsilon(3S)$ CO LDMEs are fitted to the ATLAS-2013 \cite{Upsi_ATLAS2013}
data in the fusion approximation.
$M_R^\mathcal{H}$ and $R_\mathcal{H}$ are defined in Eq.~(\ref{eq:m0}).
The errors in the CO LDMEs are multiplied by $\sqrt{\chi^2/\mbox{d.o.f.}}$ if
$\chi^2/\mbox{d.o.f.}>1$ as it is done, {\it e.g.}, in Ref.~\cite{PDG}.
The results of the NLO CPM fits for the $\psi(2S)$ meson in
Refs.~\cite{SHMZCpsi,private} and for the $\Upsilon(3S)$ meson in
Ref.~\cite{GWZUpsi} are listed for comparison.}
\label{tab:NMEs}
\end{table}
\end{turnpage}

Here and in the following, the theoretical-error bands are evaluated by
combining the scale variations and the LDME errors in quadrature.
The latter include the simultaneous variations of
$\left\langle{\cal O^H}\left[{}^1\!S_0^{(8)}\right]\right\rangle$ and
$\left\langle{\cal O^H}\left[{}^3\!P_0^{(8)}\right]\right\rangle$ in compliance
with their positivity and Eq.~(\ref{eq:m0}).
The LO KMR unPDFs \cite{KMR} adopted here are uniquely fixed by the underlying
collinear PDFs \cite{MSTW_2008}, and we neglect this source of theoretical
uncertainty.
Given the present theoretical uncertainties in the CO LDMEs, the 
hadroproduction of heavy quarkonia does not yet provide a useful laboratory to
constrain the proton unPDFs.
Precision data of the proton structure functions in deeply inelastic scattering
\cite{Hautmann:2013tba} or of the associated hadroproduction of electroweak
gauge bosons and jets \cite{Dooling:2014kia} are much more powerful in this
respect.

In the PRA, the divergent behavior of the $p_T$ distribution at small $p_T$
values is regularized by the Sudakov form factor in the KMR \cite{KMR} unPDF,
which opens the possibility to include small-$p_T$ data in the fit.
However, as is clear from Figs.~\ref{fig:ratio} and \ref{fig:CDF2009}, our
present LO-plus-LL analysis has the largest scale uncertainty in the
small-$p_T$ region, reaching a factor of 2 in the first $p_T$ bin.
Under the influence of the small-$p_T$ data, our fit slightly prefers the
${}^3\!P_J^{(8)}$ contribution, which is actually included in
Fig.~\ref{fig:CDF2009}, over the ${}^1\!S_0^{(8)}$ one. 
However, this finding should not be taken too seriously.

\begin{figure}
\begin{center}
\includegraphics[width=0.9\textwidth]{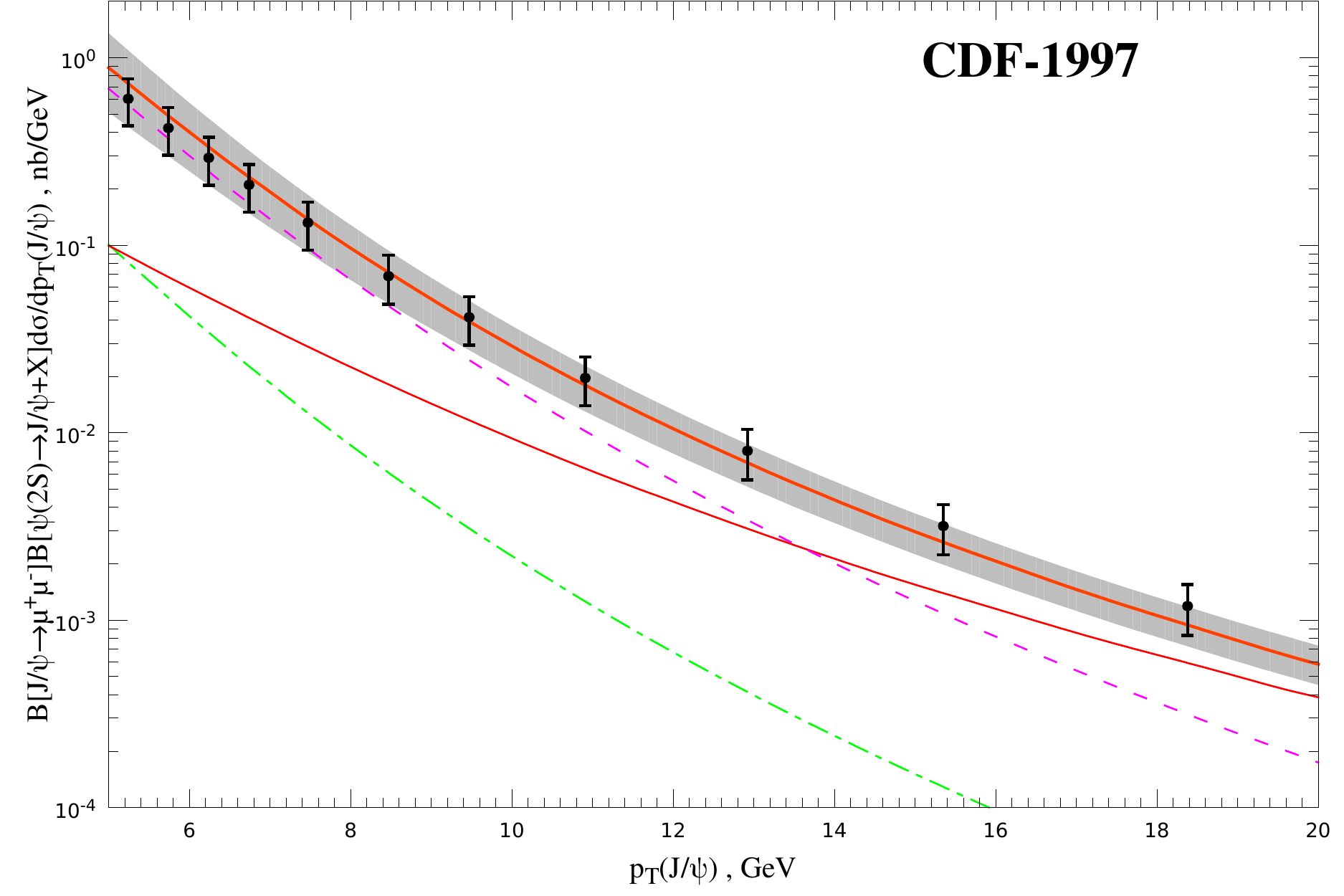}
\end{center}
\caption{\label{fig:CDF1997}%
The CDF-1997 \cite{CDF1997} data set on the $p_T^{J/\psi}$ distribution of
$J/\psi$ mesons from $\psi(2S)$ decay is compared with the predicted LO PRA
result in the fusion approximation evaluated using Eq.~(\ref{eq:pT_shift})
(thick solid orange line) and its theoretical uncertainty (shaded band).
The ${}^3\!S_1^{(1)}$ (thin dot-dashed green line),
${}^3\!S_1^{(8)}$ (thin solid red line), and
mixed ${}^1\!S_0^{(8)}$ and ${}^3\!P_J^{(8)}$ (thin dashed violet line)
contributions are shown for comparison.}
\end{figure}

The CDF Collaboration also measured the $p_T^{J/\psi}$ distribution of $J/\psi$
mesons from $\psi(2S)\to J/\psi + X$ decays at $\sqrt{S}=1.8~\mbox{TeV}$ for
$|\eta_{J/\psi}|<0.6$ in the range $5~\mbox{GeV}<p_T^{J/\psi}<20~\mbox{GeV}$
(CDF-1997) \cite{CDF1997}.
In Fig.~\ref{fig:CDF1997}, we compare these data with our LO PRA prediction
evaluated in the fusion approximation using the LDMEs determined above and with
the $p_T$ shift introduced in Eq.~(\ref{eq:pT_shift}).
We find excellent agreement within the experimental and theoretical
uncertainties, which nicely confirms the kinematic approximation underlying
Eq.~(\ref{eq:pT_shift}).

\begin{figure}
\begin{center}
\includegraphics[width=0.9\textwidth]{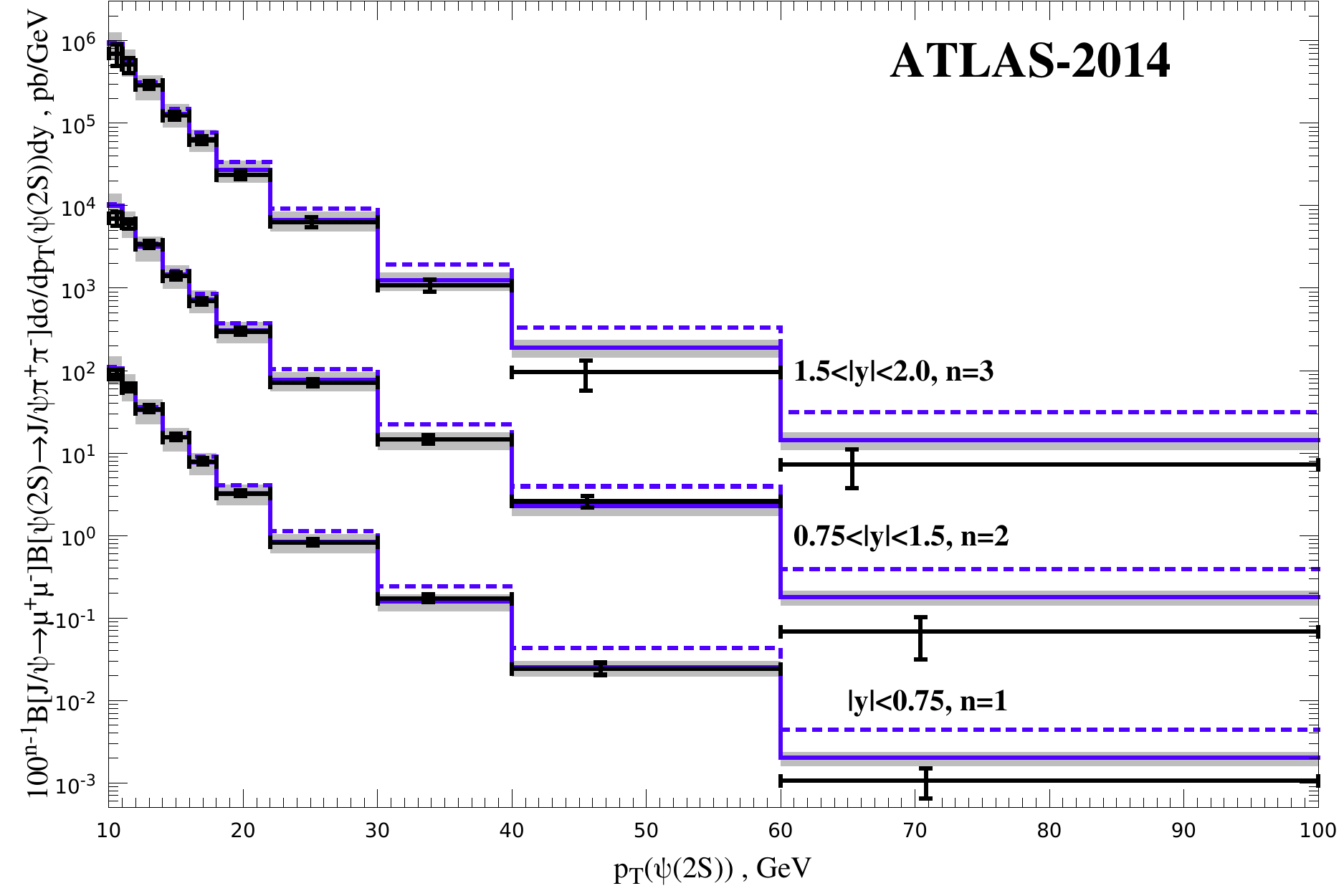}
\end{center}
\caption{\label{fig:ATLAS2014F}%
The ATLAS-2014 \cite{ATLAS2014} data sets on the $p_T$ distributions of
$\psi(2S)$ inclusive hadroproduction, multiplied by 100 for $0.75<|y|<1.5$ and
by 10\,000 for $1.5<|y|<2.0$ for better visibility, are compared with the fitted
LO PRA results in the fragmentation approximation (thick solid blue histograms)
and their theoretical uncertainties (shaded bands).
The LO PRA results in the fusion approximation (thick dashed blue histograms)
are shown for comparison.}
\end{figure}

\begin{figure}
\begin{center}
\includegraphics[width=0.9\textwidth]{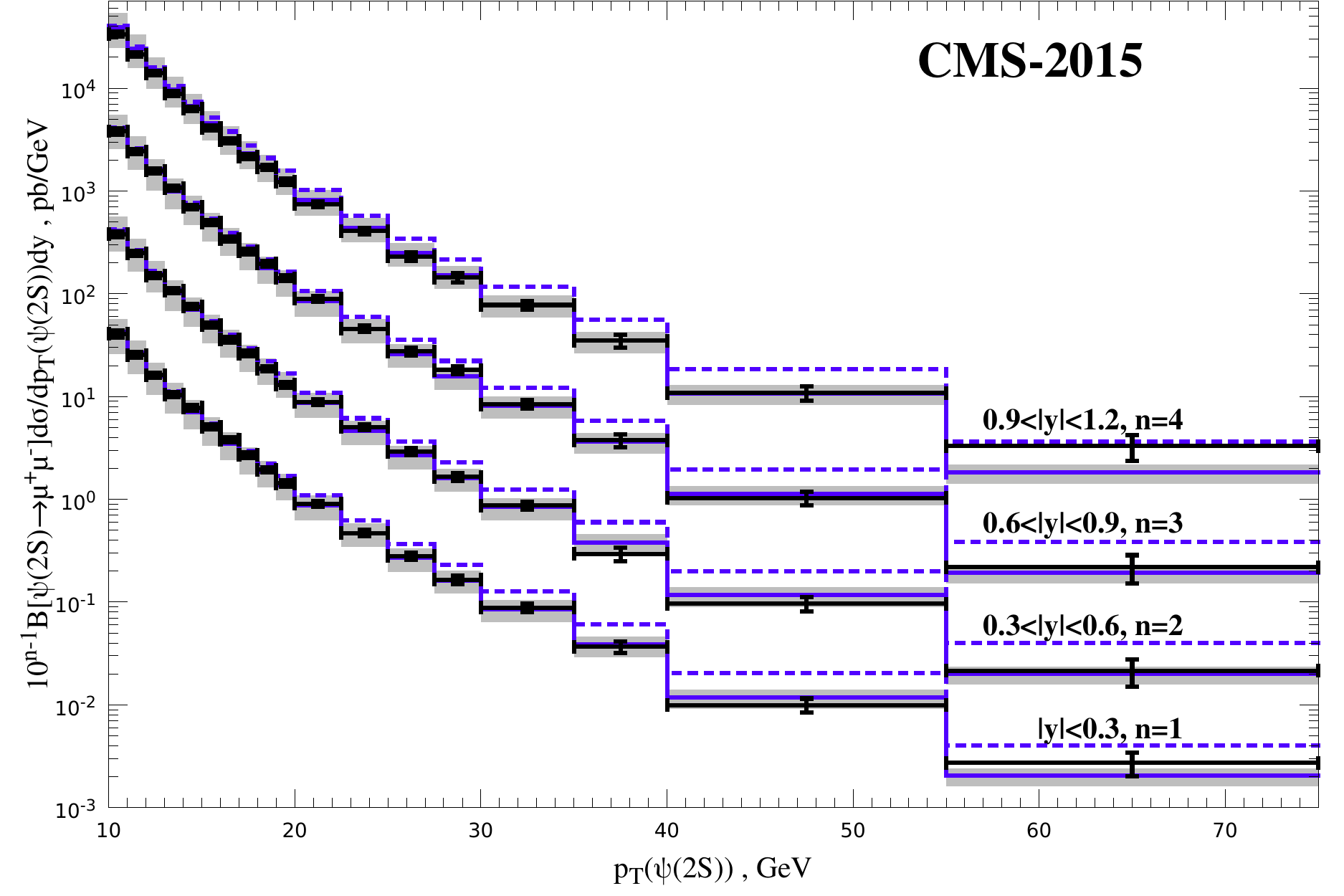}
\end{center}
\caption{\label{fig:CMS2015F}%
The CMS-2015 \cite{CMS2015} data sets on the $p_T$ distributions of $\psi(2S)$
inclusive hadroproduction, multiplied by 10 for $0.3<|y|<0.6$, by 100 for
$0.6<|y|<0.9$, and by 1\,000 for $0.9<|y|<1.2$ for better visibility, are
compared with the fitted LO PRA results in the fragmentation approximation
(thick solid blue histograms) and their theoretical uncertainties (shaded
bands).
The LO PRA results in the fusion approximation (thick dashed blue histograms)
are shown for comparison.}
\end{figure}

The ATLAS Collaboration presented their sample of
$\psi(2S)\to J/\psi+\pi^+\pi^-$ decays collected at $\sqrt{S}=7~\mbox{TeV}$
as distributions in $p_T$ and $p_T^{J/\psi}$ in the range
$10~\mbox{GeV}<p_T,p_T^{J/\psi}<100~\mbox{GeV}$ for three bins in
$|y|$ and $|y_{J/\psi}|$, respectively (ATLAS-2014) \cite{ATLAS2014}.
The CMS Collaboration measured the $p_T$ distribution of $\psi(2S)$
mesons at $\sqrt{S}=7~\mbox{TeV}$ for 4 bins in $|y|$ in the range
$10~\mbox{GeV}<p_T<75~\mbox{GeV}$ by reconstructing their
$\psi(2S)\to\mu^+\mu^-$ decays (CMS-2015) \cite{CMS2015}.
The ATLAS-2014 and CMS-2015 data may be well described in the fusion
approximation with the corresponding LDME set determined above in the lower
$p_T$ range, below 30~GeV say.
On the other hand, this approximation badly fails for the largest $p_T$ values
probed by these data.
Since the fragmentation approximation as introduced in Sec.~\ref{sec:basicPRA}
only affects the ${}^3\!S_1^{(8)}$ contribution, which is suppressed for small
values of $p_T$, as may be seen from
Figs.~\ref{fig:CDF2009} and \ref{fig:CDF1997}, it should be appropriate for the
ATLAS-2014 and CMS-2015 data, which set on at $p_T=10~\mbox{GeV}$.
In fact, our joint LO PRA fit in the fragmentation approximation to the
double-differential cross sections $d^2\sigma/(dp_T\,dy)$ measured by ATLAS
\cite{ATLAS2014} and CMS \cite{CMS2015} yield an excellent description of these
data, with $\chi^2/\mbox{d.o.f.}=1.1$, which is reflected by
Figs.~\ref{fig:ATLAS2014F} and \ref{fig:CMS2015F}, respectively.
The fit results are listed in Table~\ref{tab:NMEs};
they are in the same ball park as those extracted from the CDF-2009 data in the
fusion approximation.

For comparison, we quote in Table~\ref{tab:NMEs} also the values of the LDMEs
recently obtained through NLO CPM fits in
Refs.~\cite{GWZUpsi,SHMZCpsi,private}.\footnote{%
The fit results of Ref.~\cite{private} were used for theoretical predictions
included in Ref.~\cite{Aaij:2014qea}.}
A comparison with NLO CPM results is justified because the LO PRA approximation
captures important classes of corrections that lie beyond the LO CPM treatment.
The lack of discriminating power of the hadroproduction yield was also
experienced in Refs.~\cite{GWZUpsi,SHMZCpsi}.
By contrast, the fit in Ref.~\cite{private} included orthogonal information
from photoproduction and could so separately fix the values of
$\left\langle{\cal O}^{\psi(2S)}\left[{}^1\!S_0^{(8)}\right]\right\rangle$ and
$\left\langle{\cal O}^{\psi(2S)}\left[{}^3\!P_0^{(8)}\right]\right\rangle$,
which are combined assuming $R_{\psi(2S)}=23.0$ as in our LO PRA fit in the
fusion approximation to give the value of $M_R^{\psi(2S)}$ specified in
Table~\ref{tab:NMEs}.
The small difference between the values of
$\left\langle{\cal O}^{\psi(2S)}\left[{}^3\!S_1^{(1)}\right]\right\rangle$
extracted in Refs.~\cite{EitchenQuigg,BKL} is irrelevant for this comparison
because the ${}^3\!S_1^{(1)}$ contribution is greatly suppressed.
We observe from Table~\ref{tab:NMEs} that the NLO CPM fit results
\cite{GWZUpsi,SHMZCpsi,private} are comparable with the LO PRA ones.

Figures~\ref{fig:ATLAS2014F} and \ref{fig:CMS2015F} also contain the LO PRA
predictions evaluated in the fusion approximation using the respective LDME set
from Table~\ref{tab:NMEs}.
As anticipated above, these predictions usefully describe the ATLAS-2014 and
CMS-2015 data for $p_T\alt30$~GeV.
On the other hand, they greatly overshoot the data and their LO PRA description
in the fragmentation approximation at large $p_T$ values.
We conclude that the fusion and fragmentation approximations are consistent
in the lower $p_T$ range, and that the fragmentation corrections are very
important in the upper $p_T$ range.

\begin{figure}
\begin{center}
\includegraphics[width=0.9\textwidth]{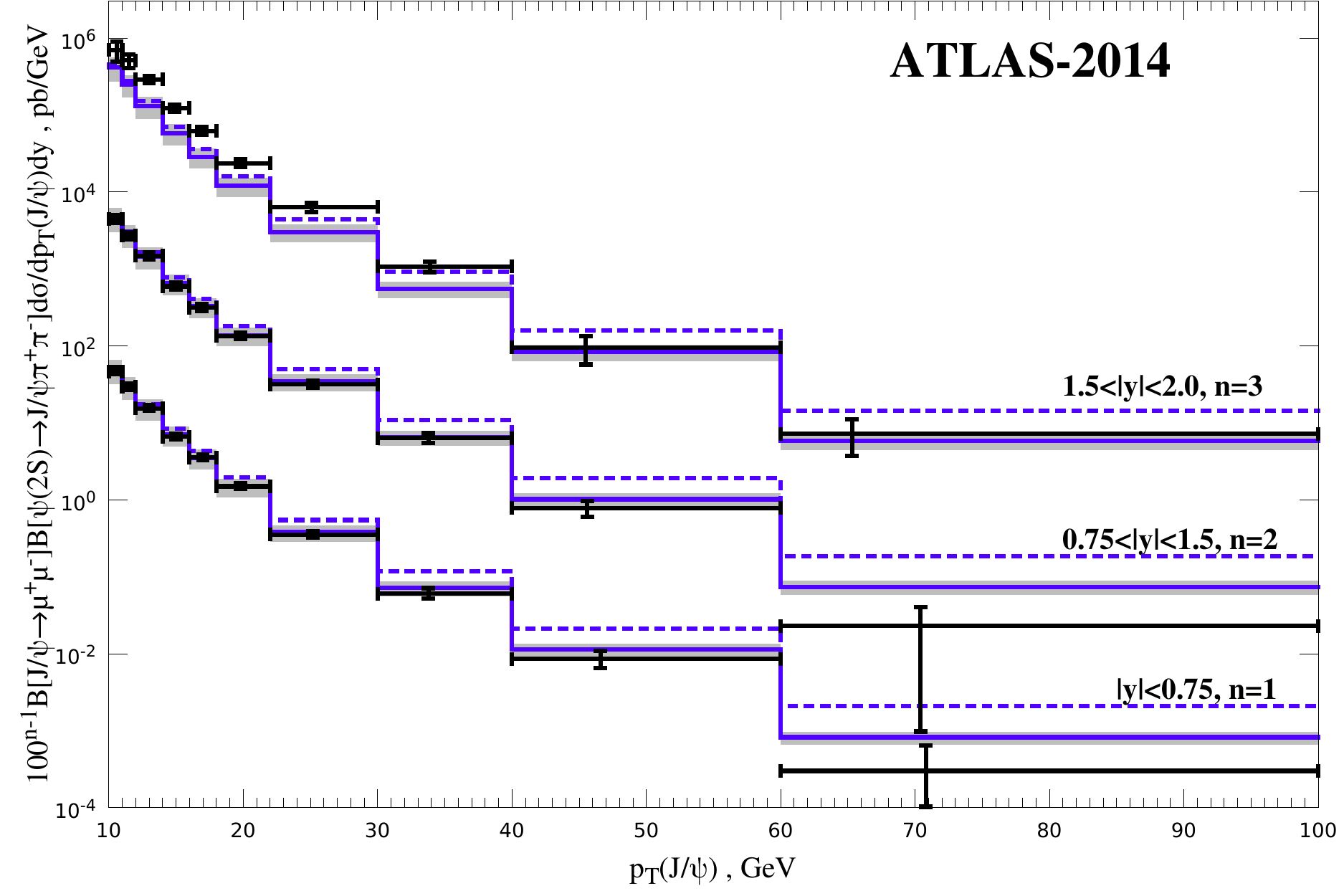}
\end{center}
\caption{\label{fig:ATLAS2014F_JPS}%
The ATLAS-2014 \cite{ATLAS2014} data sets on the $p_T^{J/\psi}$ distributions
of $J/\psi$ mesons from $\psi(2S)$ decay, multiplied by 100 for
$0.75<|y_{J/\psi}|<1.5$ and by 10\,000 for $1.5<|y_{J/\psi}|<2.0$ for better
visibility, are compared with the predicted LO PRA results in the fragmentation
approximation evaluated using Eq.~(\ref{eq:pT_shift}) (thick solid blue
histograms) and their theoretical uncertainties (shaded bands).
The LO PRA results in the fusion approximation (thick dashed blue histograms)
are shown for comparison.}
\end{figure}

In Fig.~\ref{fig:ATLAS2014F_JPS}, we compare the $p_T^{J/\psi}$ distributions
of the $J/\psi$ mesons from $\psi(2S)$ decays measured by ATLAS
\cite{ATLAS2014} in the three $|y_{J/\psi}|$ bins with our LO PRA predictions
evaluated in the fragmentation approximation with the corresponding LDMEs in
Table~\ref{tab:NMEs} and the $p_T$ shift in Eq.~(\ref{eq:pT_shift}).
For comparison, we also present the corresponding results in the fusion
approximation.
Except for the most forward $|y_{J/\psi}|$ bin, we encounter a similar
qualitative picture as in Fig.~\ref{fig:ATLAS2014F} for the $p_T$ distribution
of the ATLAS-2014 data, which is typically a factor of 2 larger.
In fact, the fragmentation approximation nicely describes the data in the
entire $p_T^{J/\psi}$ range and is consistent with the fusion approximation
in the lower $p_T^{J/\psi}$ range.
As in the case of the CDF-1997 data in Fig.~\ref{fig:CDF1997}, the kinematic
approximation in Eq.~(\ref{eq:pT_shift}) proves to be sufficiently accurate at
the LO PRA level, at least for central $y_{J/\psi}$ values.

\begin{figure}
\begin{center}
\includegraphics[width=0.9\textwidth]{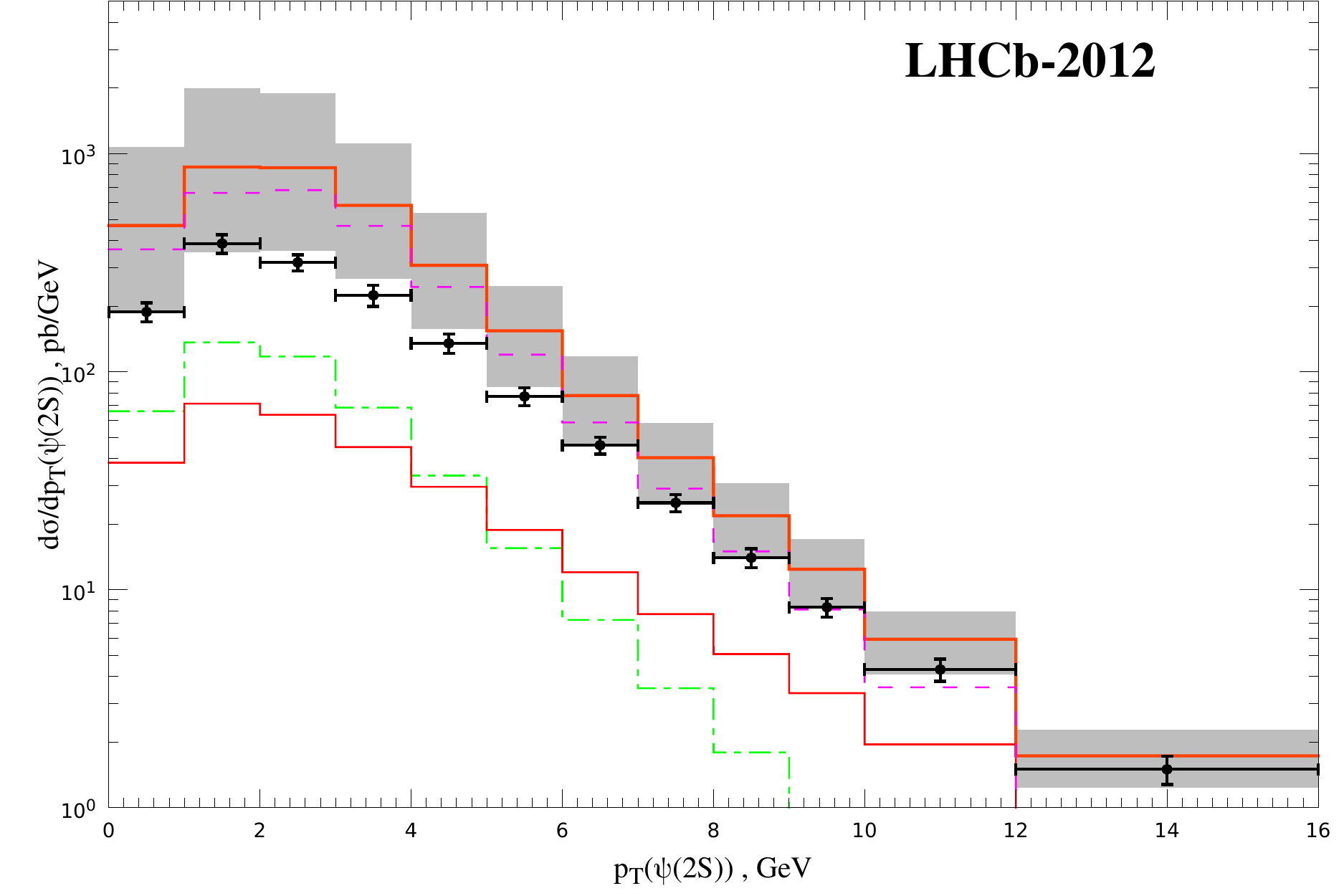}
\end{center}
\caption{\label{fig:LHCb2012}%
The LHCb-2012 \cite{LHCb2012} data set on the $p_T$ distribution of $\psi(2S)$
inclusive hadroproduction is compared with the predicted LO PRA result in the
fusion approximation (thick solid orange histogram) and its theoretical
uncertainty (shaded band).
The ${}^3\!S_1^{(1)}$ (thin dot-dashed green histogram),
${}^3\!S_1^{(8)}$ (thin solid red histogram), and
mixed ${}^1\!S_0^{(8)}$ and ${}^3\!P_J^{(8)}$ (thin dashed violet histogram)
contributions are shown for comparison.}
\end{figure}

The above comparisons were performed for measurements at central rapidities.
This kinematic region is most suitable for the application of the PRA, since
most of the initial-state radiation can be considered as highly separated in
rapidity.
The LHCb Collaboration measured the $\psi(2S)$ $p_T$ distribution for
$p_T<16~\mbox{GeV}$ at $\sqrt{S}=7~\mbox{TeV}$ in the forward region
$2.0<y<4.5$ (LHCb-2012) \cite{LHCb2012}.
In Fig.~\ref{fig:LHCb2012}, we compare this measurement with our LO PRA
predictions in the fusion approximation.
We find that the LHCb-2012 data mostly lie at the lower edge of the theoretical
error band.
We hence conclude that the LO PRA approximation with both initial-state gluons
being Reggeized is less appropriate for this kinematic region.

\begin{figure}
\begin{center}
\includegraphics[width=0.7\textwidth]{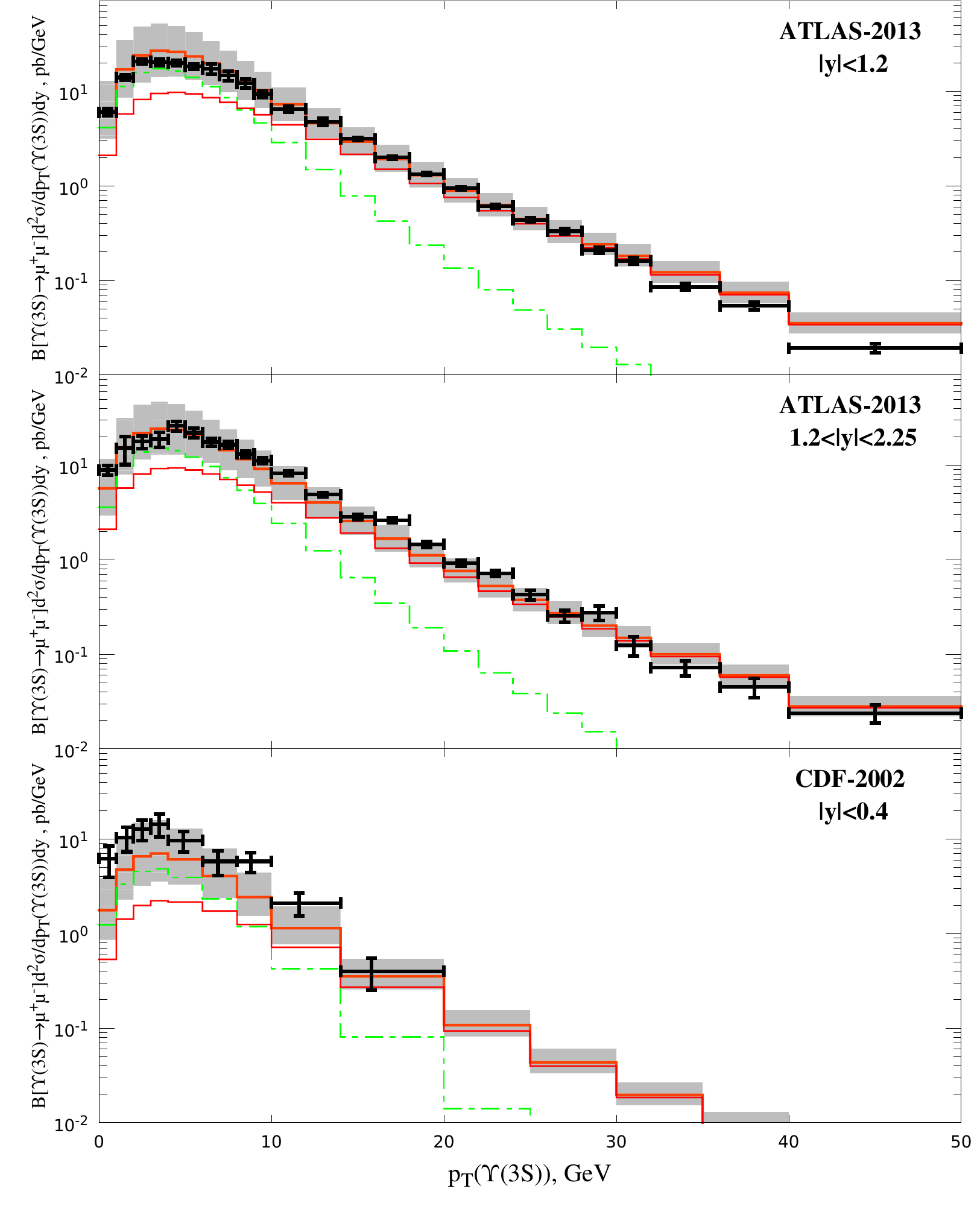}
\end{center}
\caption{\label{fig:ATLAS2013}%
The ATLAS-2013 \cite{Upsi_ATLAS2013} (upmost and center panels) and CDF-2002
\cite{Upsi_CDF2002} (downmost panel) data sets on the $p_T$ distributions of
$\Upsilon(3S)$ inclusive hadroproduction are compared with the fitted and
predicted LO PRA results in the fusion approximation (thick solid orange
histograms) and their theoretical uncertainties (shaded bands), respectively.
The ${}^3\!S_1^{(1)}$ (thin dot-dashed green histograms) and ${}^3\!S_1^{(8)}$
(thin solid red histograms) contributions are shown for comparison.
The mixed ${}^1\!S_0^{(8)}$ and ${}^3\!P_J^{(8)}$ contribution does not
contribute due to $M_R^{\Upsilon(3S)}=0$ in Table~\ref{tab:NMEs}.}
\end{figure}

We now turn to the unpolarized $\Upsilon(3S)$ yield.
The ATLAS Collaboration measured the $p_T$ distribution of $\Upsilon(3S)$
mesons at $\sqrt{S}=7~\mbox{TeV}$ in two different $|y|$ bins in the range
$p_T<50~\mbox{GeV}$ by reconstructing their $\Upsilon(3S)\to\mu^+\mu^-$ decays
(ATLAS-2013) \cite{Upsi_ATLAS2013}.
In view of $m_{\Upsilon(3S)}=10.123$~GeV, the fusion approximation is certainly
appropriate here.
Our LO PRA fit to the ATLAS-2013 data in the bins $|y|<1.2$ and
$1.2<|y|<2.5$ yields $\chi^2/\mbox{d.o.f.}=9.7$ and is presented in the upmost
and center panels of Fig.~\ref{fig:ATLAS2013}, respectively.
The hierarchy of the various contributions in the $\Upsilon(3S)$ case is
completely different from the $\psi(2S)$ case.
While the ${}^3\!S_1^{(1)}$ contribution is almost negligible in the $\psi(2S)$
case, it dominates for small $p_T$ values in the $\Upsilon(3S)$ case, leaving
little room for the ${}^1\!S_0^{(8)}$ and ${}^3\!P_J^{(8)}$ contributions.

The fit values of the CO LDMEs are listed in Table~\ref{tab:NMEs};
the ratio $R_{\Upsilon(3S)}(p_T)$ defined in Eq.~(\ref{eq:R_def}) is again
approximately constant, namely $R_{\Upsilon(3S)}=22.1\pm0.7$.
The analogous values in Ref.~\cite{SNSbottom} are slightly different because
they were obtained using $m_b=M_{\Upsilon(1S)}/2$ rather than
$m_b=M_{\Upsilon(3S)}/2$, the choice used here.
For comparison, the values of the $\Upsilon(3S)$ CO LDMEs extracted in
Ref.~\cite{GWZUpsi} are also quoted in Table~\ref{tab:NMEs}.
They are compatible with our results.

\begin{figure}
\begin{center}
\includegraphics[width=0.7\textwidth]{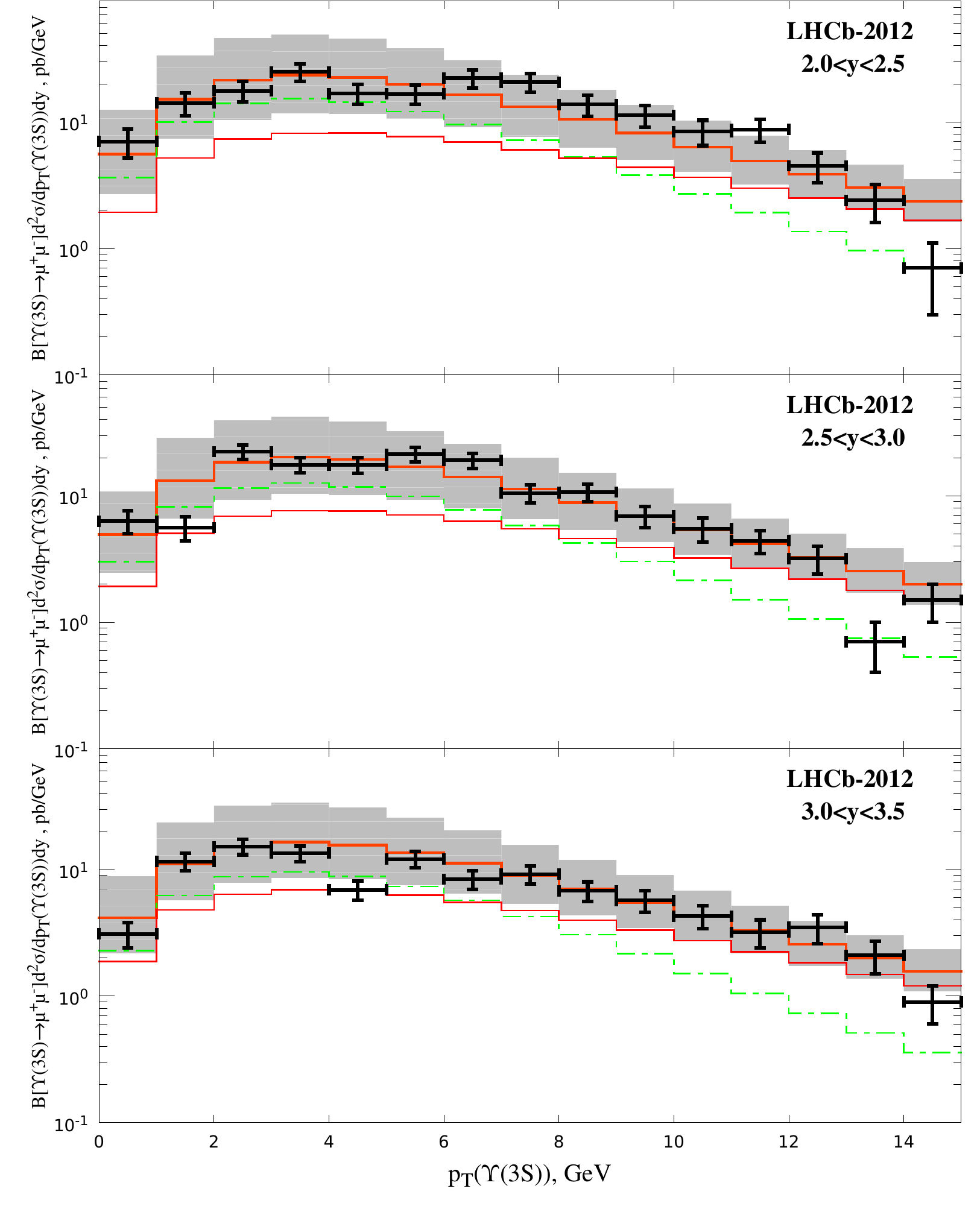}
\end{center}
\caption{\label{fig:LHCb2012_1}%
The LHCb-2012 \cite{Upsi_LHCb2012} data sets on the $p_T$ distributions of
$\Upsilon(3S)$ inclusive hadroproduction are compared with the predicted LO PRA
results in the fusion approximation (thick solid orange histograms) and their
theoretical uncertainties (shaded bands).
The ${}^3\!S_1^{(1)}$ (thin dot-dashed green histograms) and ${}^3\!S_1^{(8)}$
(thin solid red histograms) contributions are shown for comparison.
The mixed ${}^1\!S_0^{(8)}$ and ${}^3\!P_J^{(8)}$ contribution does not
contribute due to $M_R^{\Upsilon(3S)}=0$ in Table~\ref{tab:NMEs}.}
\end{figure}

\begin{figure}
\begin{center}
\includegraphics[width=0.7\textwidth]{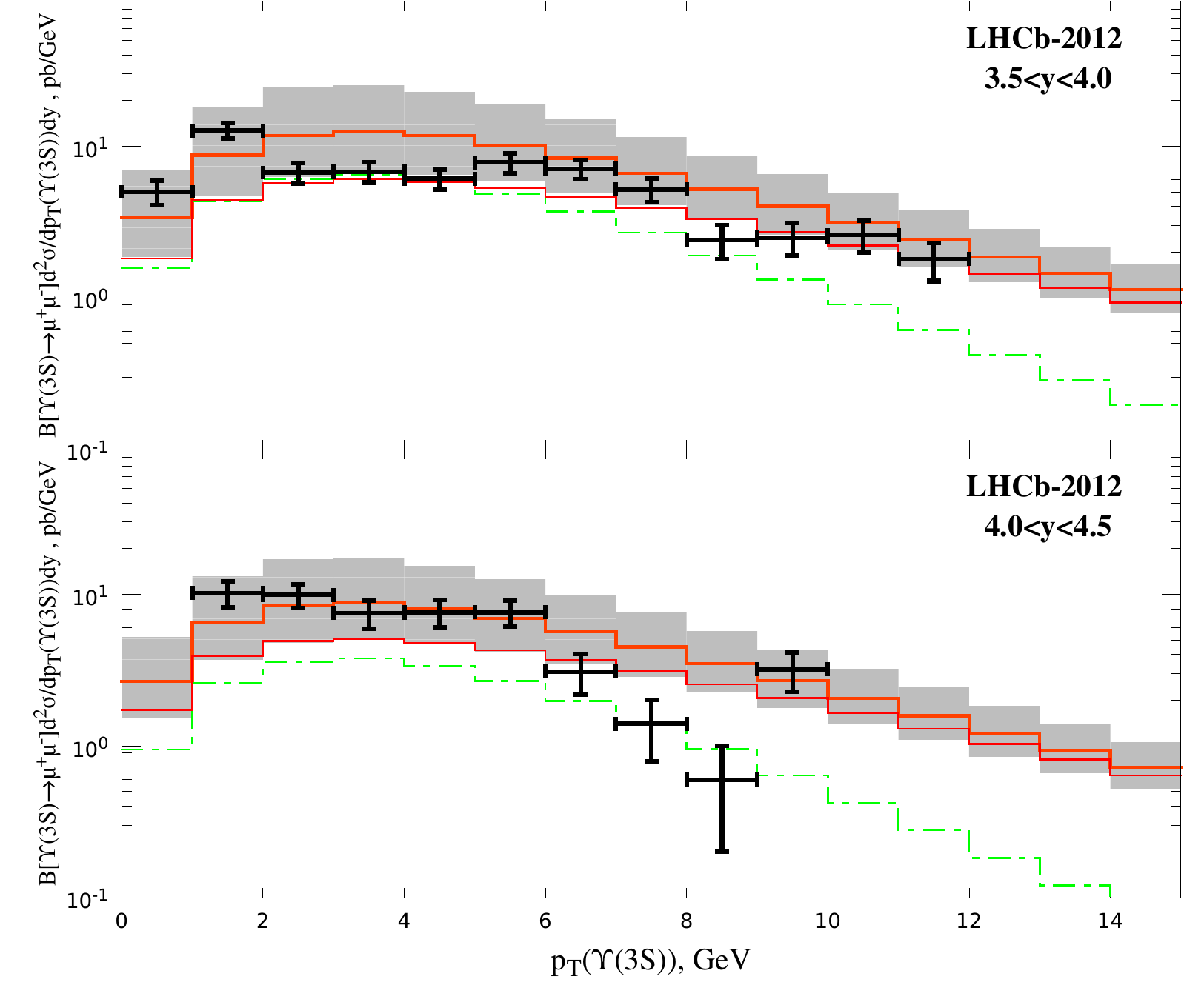}
\end{center}
\caption{\label{fig:LHCb2012_2}%
Figure~\ref{fig:LHCb2012_1} continued.}
\end{figure}

$\Upsilon(3S)$ $p_T$ distributions were also measured by the CDF Collaboration
at $\sqrt{S}=1.8~\mbox{GeV}$ for $|y|<0.4$ (CDF-2002) \cite{Upsi_CDF2002} and
by the LHCb Collaboration at $\sqrt{S}=7~\mbox{GeV}$ in five $y$ bins
(LHCb-2012a) \cite{Upsi_LHCb2012}.
These data are confronted with our LO PRA predictions in the downmost panel of
Fig.~\ref{fig:ATLAS2013} and in Figs.~\ref{fig:LHCb2012_1} and
\ref{fig:LHCb2012_2}, respectively.
The CDF-2002 data tend lie at the upper edge of our theoretical error band,
while the LHCb-2012a data exhibit nice agreement, with a few exceptions, which
appear to be runaway data points.
At this point, the question naturally arises why LO PRA works at large $y$
values for $\Upsilon(3S)$ in Figs.~\ref{fig:LHCb2012_1} and
\ref{fig:LHCb2012_2}, while it fails for $\psi(2S)$ in Fig.~\ref{fig:CDF1997}.
A possible explanation for this may be related to the fact that the 
$\Upsilon(3S)$ yield is dominated by the ${}^3\!S_1^{(1)}$ contribution at
small $p_T$ values, while the $\psi(2S)$ yield is almost exhausted by the
${}^1\!S_0^{(8)}$ and ${}^3\!P_J^{(8)}$ contributions.
In fact, the CO states could be partly destroyed by soft- or Glauber-gluon
exchanges with other partons populating the forward region, while the CS state
survives.
We, therefore, propose a more detailed study of the $y$ dependencies of the
$\psi(2S)$ and $\Upsilon(3S)$ production cross sections as a promising test of
the $k_T$-factorization-breaking effects. 

In Ref.~\cite{Chatrchyan:2013yna}, the CMS Collaboration compare their
measurements of the unpolarized $\Upsilon(nS)$ ($n=1,2,3$) yields with
theoretical predictions obtained using the CASCADE Monte Carlo event generator
\cite{Jung:2010si}, which is based on a variant of the $k_T$ factorization
formalism \cite{kTf}.
In Ref.~\cite{Jung:2010si}, the $\Upsilon(nS)$ hadroproduction cross sections
are adopted from Ref.~\cite{Baranov:2012fb}, where they are evaluated in the CS
model.
The interplay of the lack of CO contributions, the different implementation of
$k_T$ factorization, and the inclusion of nonperturbative effects beyond the
scope of our analysis, such as parton showering, render a meaningful
comparison with our results difficult.

\subsection{Polarization parameters}
\label{sec:Polar}

We now compare the $p_T$ distributions of the polarization parameters
$\lambda_\theta$ of $\psi(2S)$ and $\Upsilon(3S)$ mesons measured in the
$s$-channel helicity frame at the Tevatron and the LHC with our LO PRA
predictions.
As already pointed out in Sec.~\ref{sec:intro}, the $\psi(2S)$ and
$\Upsilon(3S)$ mesons allow for particularly pure polarization studies because
of the negligible feed-down contributions from charmonia above the $D\bar{D}$
threshold and bottomonia above the $B\bar{B}$ threshold, respectively.

In the $\psi(2S)$ case, we consider the CDF measurement at
$\sqrt{S}=1.96~\mbox{TeV}$ in the rapidity bin $|y|<0.6$ (CDF-2007)
\cite{CDFpol_psi2S} and the CMS measurement at $\sqrt{S}=7~\mbox{TeV}$ in the
bins $|y|<0.6$, $0.6<|y|<1.2$, and $1.2<|y|<1.5$ (CMS-2012) \cite{CMSpol}.
In the $\Upsilon(3S)$ case, we consider the CDF measurement at
$\sqrt{S}=1.96~\mbox{TeV}$ in the bin $|y|<0.6$ (CDF-2012)
\cite{CDFpol} and the CMS measurement at $\sqrt{S}=7~\mbox{TeV}$ in the
bins $|y|<0.6$ and $0.6<|y|<1.2$ (CMS-2012a) \cite{Chatrchyan:2012woa}.
Our LO PRA predictions are evaluated in the fusion approximation using the
respective LDMEs in Table~\ref{tab:NMEs}.
The LDME errors dominate because the scale variations largely cancel in the
ratio in Eq.~(\ref{eq:Ltheta}).

\begin{figure}
\begin{center}
\includegraphics[width=0.9\textwidth]{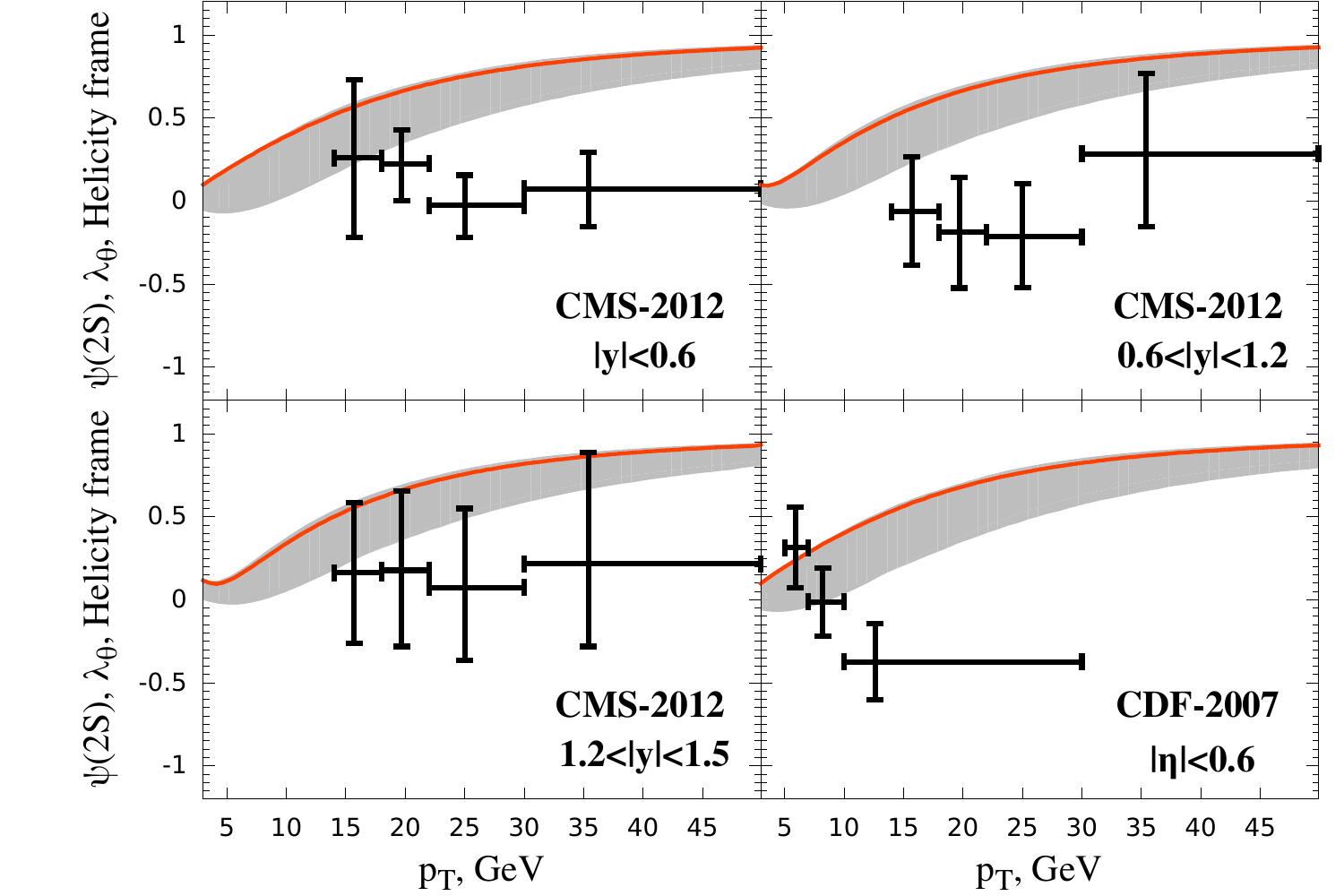}
\end{center}
\caption{\label{fig:pol_psi2S}%
The CMS-2012 \cite{CMSpol} (upper and left lower panels) and CDF-2007
\cite{CDFpol_psi2S} (right lower panel) data sets on the $p_T$ distributions of
the $\psi(2S)$ polarization parameter $\lambda_\theta$ in the $s$-channel
helicity frame are compared with the predicted LO PRA results in the
fusion approximation (thick solid orange lines) and their theoretical
uncertainties (shaded bands).}
\end{figure}

\begin{figure}
\begin{center}
\includegraphics[width=0.5\textwidth]{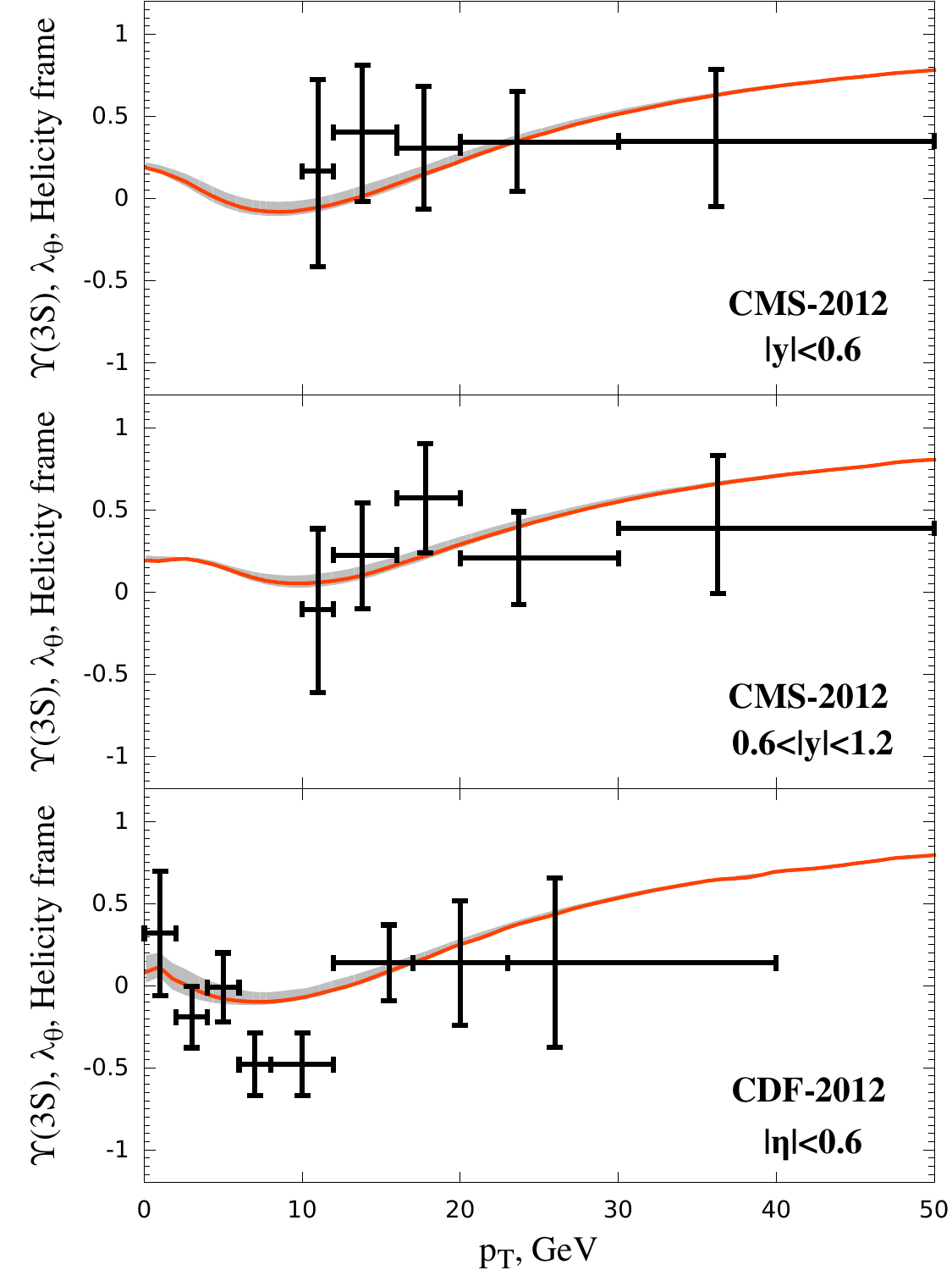}
\end{center}
\caption{\label{fig:pol_Y3S}%
The CMS-2012a \cite{Chatrchyan:2012woa} (upmost and center panels) and CDF-2012
\cite{CDFpol} (downmost panel) data sets on the $p_T$ distributions of
the $\Upsilon(3S)$ polarization parameter $\lambda_\theta$ in the $s$-channel
helicity frame are compared with the predicted LO PRA results in the
fusion approximation (thick solid orange lines) and their theoretical
uncertainties (shaded bands).}
\end{figure}

The comparisons for the $\psi(2S)$ and $\Upsilon(3S)$ mesons are shown in
Figs.~\ref{fig:pol_psi2S} and \ref{fig:pol_Y3S}, respectively.
From Fig.~\ref{fig:pol_psi2S} we observe that the LO PRA predictions tend to
overshoot the experimental data \cite{CDFpol_psi2S,CMSpol} at large $p_T$
values.
In fact, the $\psi(2S)$ mesons are predicted to be asymptotically transverse,
with $\lambda_\theta=1$, in the large-$p_T$ limit.
There, the cross section is practically saturated by the production of the
${}^3\!S_1^{(8)}$ state via an almost on-shell gluon, which passes on its
transverse polarization via the ${}^3\!S_1^{(8)}$ state to the $\psi(2S)$ meson.
We thus recover the notion charmonium polarization puzzle, which is familiar
from the CPM \cite{KBpol,SHMZCpsi}.

By contrast, in the $\Upsilon(3S)$ case featured in Fig.~\ref{fig:pol_Y3S},
there is excellent agreement between the experimental data
\cite{CDFpol,Chatrchyan:2012woa} and our LO PRA predictions, with the
exceptions of two CDF-2012 data points.
As in the $\psi(2S)$ case, the experimental data are essentially compatible
with zero polarization.
However, the $g\to b\bar{b}[{}^3\!S_1^{(8)}]$ transition does not play a
dominant role in the $p_T$ range considered.
Similar observations were made at NLO in the CPM \cite{GWZUpsi}.

\section{Conclusions}
\label{sec:Concl}

In the present paper, we studied the hadroproduction of $\psi(2S)$ and
$\Upsilon(3S)$ mesons at the Tevatron and the LHC in the NRQCD factorization
approach working at LO in the PRA.
These are particularly clean probes because the contaminations from feed-down
contributions are negligibly small.
We considered the unpolarized yields and the polarization parameter
$\lambda_\theta$ in the $s$-channel helicity frame as functions of $p_T$.
While the analytic results for the unpolarized yields are already available in
the literature \cite{KSVcharm}, we provided those for $\lambda_\theta$ here.

In the $\psi(2S)$ case, we extracted two sets of CO LDMEs, one by fitting the
CDF-2009 \cite{CDF2009} data in the fusion approximation and one by jointly
fitting the ATLAS-2014 \cite{ATLAS2014} and CMS-2015 \cite{CMS2015} data in the
fragmentation approximation.
We found that the fusion approximation usefully describes the ATLAS-2014 and
CMS-2015 data in the lower $p_T$ range, for $p_T\alt30~\mbox{GeV}$, while the
fragmentation approximation is indispensable for larger $p_T$ values.
However, we encountered limitations of the PRA at LO in describing the
LHCb-2012 \cite{LHCb2012} measurement in the forward direction.
We also verified that that the simple kinematic approximation in
Eq.~(\ref{eq:pT_shift}) leads to a satisfactory description of the CDF-1997
\cite{CDF1997} and ATLAS-2014 \cite{ATLAS2014} data on the $p_T^{J/\psi}$
distributions of the $J/\psi$ mesons from $\psi(2S)$ decay.
By confronting the CDF-2007 \cite{CDFpol_psi2S} and CMS-2012 \cite{CMSpol} data
on $\lambda_\theta$ with our predictions in the fusion approximation, we found
that the charmonium polarization puzzle, which is familiar from the CPM both
at LO \cite{POLfrms,BKL,KniehlLee} and NLO \cite{KBpol}, persists at LO in the
PRA.

The situation is very different in the $\Upsilon(3S)$ case.
Thanks to $M_{\Upsilon(3S)}\gg M_{\psi(2S)}$, the fusion approximation is quite
appropriate in the $p_T$ range experimentally accessed so far, and the PRA at
LO usefully works also in the forward direction.
In fact, the set of CO LDMEs that we fitted to the ATLAS-2013
\cite{Upsi_ATLAS2013} data yield a nice description of the LHCb-2012a
\cite{Upsi_LHCb2012} data, albeit the one of the CDF-2002 \cite{Upsi_CDF2002}
data is marginal.
Furthermore, CDF-2012 \cite{CDFpol} and CMS-2012 \cite{CMSpol} data on
$\lambda_\theta$ agree very well with our LO PRA predictions, which we
attributed to the subdominant role of the $g\to b\bar{b}[{}^3\!S_1^{(8)}]$
transition.

In conclusion, the PRA once again proved to be a powerful tool for the
theoretical description of QCD processes in the high-energy limit.
It allows one to achieve useful descriptions of experimental data already at
LO in cases when one needs to go to NLO or perform resummations in the CPM.
This is in line with our previous studies in the PRA,
applied to the production of charmonia
\cite{Frag_YaF,KSVcharm,Frag_PRD,SVYadFiz,SNScharm}, bottomonia
\cite{SVYadFiz,KSVbottom,SNSbottom}, $D$ mesons \cite{OpenCharm}, $B$ mesons
\cite{Bmesons}, dijets \cite{dijets}, bottom-flavored jets \cite{bjets},
Drell-Yan lepton pairs \cite{DY}, monojets, and prompt photons
\cite{PPSJ,PPJHera}.
On the other hand, the PRA at LO fails to solve the charmonium polarization
puzzle.
Our study indicates that the latter is an intrinsic problem of NRQCD
factorization in the final state and rather insensitive to the treatment of
gluonic initial-state radiation.

\section*{Acknowledgments}

The authors are grateful to M.~Butensch\"on for useful discussions and to
C.~Louren\c{c}o for a clarifying communication regarding the CMS-2015 data
\cite{CMS2015}.
The work of B.~A.~K. was supported in part by the German Federal Ministry for
Education and Research BMBF through Grant No.~05H2015.
The work of M.~A.~N. and V.~A.~S. was supported in part by the Russian
Foundation for Basic Research through Grant No.~14-02-00021 and by the
Competitiveness Enhancement Program of Samara University for 2013--2020.
The work of M.~A.~N. was supported in part by the German Academic Exchange
Service DAAD and by the Ministry of Education and Science of the Russian
Federation through ``Michail Lomonosov'' Grant No.~A/14/73130 and through
Grant No.~1394. 



\end{document}